\documentclass[reprint,amsmath,amssymb,aps,prc,nofootinbib,superscriptaddress]{revtex4-1}
\usepackage{bm}
\usepackage{graphicx}
\usepackage{dcolumn}
\usepackage{color}
\newcommand{\be}{\begin{equation}}
\newcommand{\ee}{\end{equation}}
\newcommand{\bea}{\begin{eqnarray}}
\newcommand{\eea}{\end{eqnarray}}
\newcommand{\bfr}{\mbox{\boldmath $r$}}

\newcommand{\sbfr}{\mbox{\scriptsize\boldmath $r$}}
\newcommand{\sbfrp}{\mbox{\scriptsize\boldmath $r'$}}

\newcommand{\bfs}{\mbox{\boldmath $s$}}

\newcommand{\bfl}{\mbox{\boldmath $l$}}
\newcommand{\bfj}{\mbox{\boldmath $j$}}
\newcommand{\bfbj}{\mbox{\boldmath $J$}}
\newcommand{\bfbt}{\mbox{\boldmath $T$}}

\newcommand{\bfbq}{\mbox{\boldmath $Q$}}

\newcommand{\bftau}{\mbox{\boldmath $\tau$}}
\newcommand{\bftaup}{\mbox{\boldmath $\tau'$}}

\newcommand{\mbss}[1]{_{\mbox{\scriptsize #1}}}

\newcommand{\mbsu}[1]{\mbox{\scriptsize #1}}

\newcommand{\vphu}{\vphantom{*}}
\newcommand{\vphd}{\vphantom{1}}
\newcommand{\hfm}{\hphantom{-}}
\newcommand{\hfz}{\hphantom{0}}
\newcommand{\bfsigma}{\mbox{\boldmath $\sigma$}}
\newcommand{\bfsigmap}{\mbox{\boldmath $\sigma'$}}

\newcommand{\ve}{\varepsilon}

\newcommand{\half}{{\textstyle\frac{1}{2}}}
\newcommand{\ffrac}[2]{{\textstyle\frac{#1}{#2}}}

\begin{document}

\title{Low-energy M1 excitations in $^{208}$Pb\\
and the spin channel of the Skyrme energy-density functional
}

\author{V. Tselyaev}
\affiliation{St. Petersburg State University, St. Petersburg, 199034, Russia}
\email{tselyaev@mail.ru}
\author{N. Lyutorovich}
\affiliation{St. Petersburg State University, St. Petersburg, 199034, Russia}
\author{J. Speth}
\affiliation{Institut f\"ur Kernphysik, Forschungszentrum J\"ulich, D-52425 J\"ulich, Germany}
\author{P.-G. Reinhard}
\affiliation{Institut f\"ur Theoretische Physik II, Universit\"at Erlangen-N\"urnberg,
D-91058 Erlangen, Germany}
\author{D. Smirnov}
\affiliation{St. Petersburg State University, St. Petersburg, 199034, Russia}
\date{\today}

\begin{abstract}
We investigate the spin dependent part of the Skyrme energy-density
functional, in particular its impact on the residual particle-hole
interaction in self-consistent calculations of excitations.  Test
cases are the low-energy M1 excitations in $^{208}$Pb treated within
the self-consistent random-phase approximation based on the Skyrme
energy-density functional. We investigate different parametrizations
of the functionals to find out which parameters of the functional have
strongest correlations with M1 properties. We explore a simple method
of the modification of the spin-related parameters which delivers a
better description of M1 excitations while basically maintaining the
good description of ground state properties.
\end{abstract}


\maketitle

\section{Introduction}
\label{sec:Intr}

The aim of this paper is to explore the description of nuclear
magnetic excitations by an energy-density functional (EDF) of Skyrme
type \cite{Bender_2003} taking the low-lying magnetic dipole
  (M1) excitations in $^{208}$Pb as test case. The random phase
approximation (RPA) and its various extensions is the most often used
method for the investigation of nuclear excitation spectra. It takes
as input data single-particle ($sp$) energies, $sp$ wave functions
and a particle-hole ($ph$) residual interaction.  Early calculations
as, e.g., Migdal's Theory of Finite Fermi Systems (TFFS, see
Refs.~\cite{Migdal_1967,Ring_1973PLB,Borzov_1984}) started with an
effective single-particle model whose parameters are adjusted to
experimental $sp$-properties and used (in nearly all numerical
applications) a density-dependent zero-range $ph$-interaction.  It
requires only a few parameters, coined Landau-Migdal (LM) parameters,
which are adjusted to electric and magnetic nuclear excitations and
which turn out to be universal in the sense that the same values apply
throughout the chart of nuclei \cite{Speth_1991}.  In self-consistent
nuclear models, one obtains the $sp$-properties as well as the
$ph$-interaction from one and the same effective Hamiltonian, or EDF
respectively.  The parameters of the Skyrme EDF are primarily adjusted
to bulk properties of the nuclear ground state.  An appropriate $ph$
residual interaction is not a priori guaranteed.  For example, the
first realistic Skyrme parametrizations \cite{VaBr,Beiner_1975} had an
incompressibility of the order of 350 MeV and produced therefore the
breathing mode in $^{208}$Pb at around 17 MeV (which was off by 3 MeV
from the experimental value measured some years later).  Including
data specific to excitations, one could later on develop
parametrizations which also perform well for breathing mode and
isoscalar quadrupole resonance \cite{Bartel_1982,Brack_1985}.  In
general, there is sufficient flexibility in the Skyrme EDF to
accommodate all modes with natural parity, isoscalar as well as
isovector resonances \cite{Kluepfel_2009}. The LM parameters for
natural-parity excitations derived from such Skyrme EDFs agree nicely
with long tested LM parameters of TFFS \cite{Speth_2014}.

For magnetic modes, self-consistent models as, e.g., Skyrme EDFs have
not yet reached that high level of descriptive power while TFFS has
been adapted very well also for these excitation channels. The plan
for this paper is thus to explore the chances for a better description
of magnetic modes with a Skyrme EDF exploiting yet loosely determined
aspects of the functional.  Here we let us guide from the large body
of experience gathered within the TFFS. It tells us that the spin
dependent $ph$-interaction is weak for the isoscalar part and is
strongly repulsive for the isovector part. This agrees with the
experimental findings: There are no isoscalar collective magnetic
resonances known over the whole periodic system but there exist strong
Gamow-Teller resonances in heavy nuclei which are created by the
spin-isospin dependent part of the residual interaction. We also know
from such investigations that the M1 states in $^{208}$Pb represent an
ideal test case.  Experimental data on the distribution of the M1
strength in this nucleus at the excitation energies up to 8.4~MeV are
known since the work of
\cite{Wienhard_1982,Koehler_1987,Laszewski_1988}.  Updates for the
energies below neutron separation energy were published in
\cite{Shizuma_2008}.  The observed spectrum of the low-energy M1
excitations in $^{208}$Pb consists of two marked features: an
isoscalar $1^+$ state with $E$ = 5.844~MeV and a broad isovector M1
resonance in the interval 6.6--8.1 MeV.  Strong fragmentation of the
M1 resonance was one of the reasons of the difficulties with
identification in the early experiments (see, e.g.,
Ref.~\cite{Bertsch_1981} for discussion).  Moreover, several states
which had been originally identified as M1 turned out to be E1 after
experiments with polarized photons were available.

The numerous theoretical papers devoted to the microscopic description
of M1 excitations in $^{208}$Pb can be divided into two main groups.
The first group includes the papers in which the nuclear excitations
are treated as superposition of the one-particle--one-hole
($1ph$) configurations, that is within the RPA or the Tamm-Dancoff
approximation (see, in particular, Refs.
\cite{Vergados_1971,Ring_1973PLB,Speth_1980,Borzov_1984,Migli_1991,
  Cao_2009,Vesely_2009,Nesterenko_2010,Cao_2011,Wen_2014}).  In the
papers of the second group, various versions of beyond-RPA
  approaches are used in which the RPA configuration space is
enlarged by adding the $2ph$, $1ph\otimes$phonon or two-phonons
configurations (see, e.g.,
\cite{Dehesa_1977,Kamerdzhiev_1984,Cha_1984,Khoa_1986,
  Kamerdzhiev_1989,Tselyaev_1989,Kamerdzhiev_1993a}).
Most of the earlier work as mentioned before was performed within the
TFFS. Using experimental single-particle energies as input for the
mean-field part and properly tuning the interaction parameters (LM
parameters) in the spin-spin channel, they managed to provide an
appropriate description of peaks and M1 strengths. Beyond-RPA
treatments, properly including the coupling of $1ph$ states to $2ph$ configurations,
were necessary to describe the
spectral fragmentation of the M1 resonance around 7.5
MeV \cite{Kamerdzhiev_1993a}.

Fully self-consistent RPA calculations as done in
\cite{Cao_2009,Vesely_2009,Nesterenko_2010,Cao_2011,Wen_2014} did not
yet reach that level of description. In fact, there is no published
Skyrme parametrization which can describe simultaneously position and
strength of M1 modes in $^{208}$Pb and other nuclei
\cite{Vesely_2009,Nesterenko_2010}. Already $^{208}$Pb alone seems to
pose insurmountable difficulties. It is hard to get the lower M1 peak
and the M1 resonance simultaneously at their correct energies, not to
mention a proper prediction of M1 strength.  Inappropriate strengths
of spin-orbit coupling were identified as one major source of the
problem \cite{Vesely_2009,Nesterenko_2010}.  We had applied a recently
optimized phonon-coupling model on top of self-consistent RPA
\cite{Tselyaev_2017,Tselyaev_2018} to M1 modes and, unfortunately, did
not find any improvement concerning spectral separation of low and
upper mode nor sufficiently strong fragmentation. The problem has
first to be cleared at RPA level before invoking more advanced
approaches. The first task to be solved is thus to develop a Skyrme
parametrization which describes energies and strengths of the leading
M1 modes correctly. And this is what we will attack in the present
paper, namely to work out the crucial handles in the Skyrme energy
functionals which have most impact in the M1 spectrum and to try to
tune them to deliver correct M1 spectra without spoiling the high
quality with respect to nuclear ground state observables.

The paper is organized as follows:
Section \ref{sec:method} provides the formal background of RPA, the
Skyrme functional, the magnetic operators, and the numerical scheme.
Section \ref{sec:result} discusses M1 modes in the context of Skyrme
EDFs and works out the leading mechanisms defining these modes.
In Section \ref{sec:modpar} we try a moderate readjustment of Skyrme
parameters which leads to better description of M1 modes.
The last section contains the conclusions.

\section{Formal background}
\label{sec:method}

\subsection{Summary of the RPA}
\label{sec:SumRPA}

Within the RPA one can calculate the spectrum of the excitation
energies $\omega^{\vphu}_{n}$ of the even-even nucleus
and the corresponding set of the transition amplitudes $Z^{n}_{12}$
where the numerical indices $(1, 2, 3, \ldots)$ stand for the
sets of the quantum numbers of some single-particle basis.
Generally, this basis can be arbitrary, but it is convenient
to suppose that it diagonalizes the single-particle density matrix
$\rho^{\vphu}_{12}$ and the single-particle Hamiltonian $h^{\vphu}_{12}$
which satisfy the relations $\rho^2 = \rho$ and $[\,h,\rho\,]=0$.
In this case the following equations are fulfilled
\be
h^{\vphu}_{12} = \ve^{\vphu}_{1}\delta^{\vphu}_{12}\,,
\qquad
\rho^{\vphu}_{12} = n^{\vphu}_{1}\delta^{\vphu}_{12}\,.
\label{spbas}
\ee
In what follows the indices $p$ and $h$ will be used to label
the single-particle states of the particles ($n^{\vphu}_{p} = 0$)
and holes ($n^{\vphu}_{h} = 1$) in this basis.

The RPA eigenvalue equation has the form
\be
\sum_{34} \Omega^{\mbss{RPA}}_{12,34}\,Z^{n}_{34} =
\omega^{\vphu}_n\,Z^{n}_{12}\,,
\label{rpaze}
\ee
where
\be
\Omega^{\mbss{RPA}}_{12,34} =
h^{\vphu}_{13}\,\delta^{\vphu}_{42} -
\delta^{\vphu}_{13}\,h^{\vphu}_{42}
+ \sum_{56} M^{\mbss{RPA}}_{12,56}\,{V}^{\vphu}_{56,34}\,,
\label{omrpa}
\ee
\be
M^{\mbss{RPA}}_{12,34} =
\delta^{\vphu}_{13}\,\rho^{\vphu}_{42} -
\rho^{\vphu}_{13}\,\delta^{\vphu}_{42}\,,
\label{mrpa}
\ee
$V$ is the amplitude of the residual interaction
and $M^{\mbss{RPA}}_{\vphd}$ is the metric matrix.
The matrices $\Omega^{\mbss{RPA}}_{\vphd}$ and $M^{\mbss{RPA}}_{\vphd}$
act in the ``$ph+hp$'' space.
The transition amplitudes $Z^{n}_{12}$ are normalized to
\be
\sum_{1234}
{Z}^{n*}_{12}\,M^{\mbss{RPA}}_{12,34}\,{Z}^{n'}_{34} =
\mbox{sgn}(\omega^{\vphu}_{n})\,\delta^{\vphu}_{n,\,n'}.
\label{zmz}
\ee
In the self-consistent RPA, which is supposed in the following,
the following relations are fulfilled:
\be
h^{\vphu}_{12} = \frac{\delta E[\rho]}{\delta\rho^{\vphu}_{21}}\,,\qquad
{V}^{\vphu}_{12,34} =
\frac{\delta^2 E[\rho]}
{\delta\rho^{\vphu}_{21}\,\delta\rho^{\vphu}_{34}}\,,
\label{sccond}
\ee
where $E[\rho]$ is an energy density functional.

The amplitudes $Z^{n}_{12}$ allow us to calculate the reduced probabilities
of the transitions caused by the external field operator $Q^{\alpha}_{LM}$
according to the formula
\be
B_n (\alpha L_n) = \sum_{M_n} |\langle\,{Z}^{n}\,|\,Q^{\alpha}_{L_n M_n} \rangle|^2,
\label{redprob}
\ee
where index $\alpha$ labels different kinds of the operators of the
multipolarity $L$ (in particular, $\alpha =$ m for the magnetic transitions).

\subsection{The Skyrme energy density functional}
\label{sec:EDF}

As the energy density functional $E[\rho]$ in Eqs. (\ref{sccond}) we take
the Skyrme EDF of the standard form
(see, e.g., Refs.~\cite{Dobaczewski_1995,Dobaczewski_1996}).
It can be represented as the sum of the following terms
\be
E_\mathrm{Skyrme} = E_\mathrm{kin} + E_\mathrm{int} + E_\mathrm{Coul}
\label{ESk3}
\ee
where
\be
E_\mathrm{kin} = \int d\bfr\,
\Bigl[\,\frac{\hbar^2}{2m_p} \tau_p(\bfr) + \frac{\hbar^2}{2m_n} \tau_n(\bfr)\,\Bigr]\,,
\label{Ekin}
\ee
\be
E_\mathrm{int} = \int d\bfr\,\mathcal{E}_\mathrm{int}(\bfr)\,,
\label{Eint}
\ee
\bea
E_\mathrm{Coul} &=&
  \frac{e^2}{2} \int d\bfr\,d\bfr^\prime\,
   \frac{\rho_p(\bfr)\rho_p(\bfr^\prime)}{|\bfr - \bfr^\prime|}
\nonumber\\
&-&
   \frac{3e^2}{4}\left(\frac{3}{\pi}\right)^{1/3}
   \int d\bfr \rho_p^{4/3}(\bfr)\,.
\label{ECoul}
\eea
The energy density in Eq. (\ref{Eint}) is given by
\bea
\mathcal{E}_\mathrm{int} &=& \sum_{T=0,1} \Bigl[
C_T^\rho\,\rho_T^2 +
C_T^{\rho,\alpha}\,\rho_T^2\rho_0^\alpha +
C_T^{\Delta\rho}\,\rho^{\vphu}_T\Delta\rho^{\vphu}_T
\nonumber\\
&+&
C_T^{\tau}\,\bigl(\,\rho^{\vphu}_T \tau^{\vphu}_T - \bfj_T^2\,\bigr) +
C_T^{J}\,\bigl(\,\bfbj_T^2 - \bfs^{\vphu}_{T} \!\cdot\! \bfbt^{\vphu}_{T}\,\bigr)
\nonumber\\
&+&
C_T^{\nabla J}\,\bigl(\,\rho^{\vphu}_T\nabla\!\cdot\!\bfbj^{\vphu}_T
+ \bfs^{\vphu}_{T}\!\cdot\!\nabla\!\times\!\bfj_T \,\bigr)
\nonumber\\
&+&
C_T^{s}\,\bfs_T^2 +
C_T^{s,\alpha}\,\bfs_T^2\rho_0^\alpha +
C_T^{\Delta s}\,\bfs^{\vphu}_T \Delta \bfs^{\vphu}_T\,
\Bigr]
\label{Eden}
\eea
where
$\,C_T^\rho$,
$\,C_T^{\rho,\alpha}$,
$\,C_T^{\Delta\rho}$,
$\,C_T^{\tau}$,
$\,C_T^{J}$,
$\,C_T^{\nabla J}$,
$\,C_T^{s}$,
$\,C_T^{s,\alpha}$,
$\,C_T^{\Delta s}$, and
$\,\alpha$
are the constants,
$\,\rho^{\vphu}_T$,
$\,\tau^{\vphu}_T$,
$\,\bfbj^{\vphu}_T$,
$\,\bfs^{\vphu}_{T}$,
$\,\bfbt^{\vphu}_{T}$, and
$\,\bfj^{\vphu}_T$
are the local densities and currents.
These densities and currents are divided
into two groups (see \cite{Bender_2003,Engel_1975}):
time-even
($\,\rho^{\vphu}_T$,
$\,\tau^{\vphu}_T$,
$\,\bfbj^{\vphu}_T$) and
time-odd
($\,\bfs^{\vphu}_{T}$,
$\,\bfbt^{\vphu}_{T}$,
$\,\bfj^{\vphu}_T$).
Their definition through the single-particle density matrix is given in
Appendix~\ref{app:dens}.

In the general case, if the form of the functional $E_\mathrm{int}$ is constrained
only by the conditions of the global symmetries, the $C$-constants are
the independent parameters.
Usually, they are determined by fitting the results of the Skyrme-Hartree-Fock (SHF)
and RPA calculations
to the experimental data on basic nuclear properties with taking into account
the constraints imposed by the nuclear matter properties.
However, if the Skyrme EDF, Eqs. (\ref{ESk3})--(\ref{Eden}), is derived
within the Hartree-Fock approximation from the many-body Hamiltonian containing
two-body velocity and density dependent zero-range interaction, the number of
the independent $C$-constants decreases. In this case 18 $C$-constants
in Eq.~(\ref{Eden}) are expressed through 10 Skyrme-force parameters
$t_0$, $x_0$, $t_1$, $x_1$, $t_2$, $x_2$, $t_3$, $x_3$, $W_0$, and $x_W$
(see, e.g., \cite{Bender_2003}).
The respective formulas are given in Appendix~\ref{app:param}.

Different bias in choosing the data and steady growth of information
on exotic nuclei has lead to a great variety of parametrizations.  In
order to keep the present survey sufficiently general, we consider a
large set of 30 different parametrizations of the Skyrme EDF: SIII
\cite{Beiner_1975}, SGII \cite{vanGiai_1981}, SkM$^*$
\cite{Bartel_1982,Brack_1985}, RATP \cite{Rayet_1982}, T5 and T6
\cite{Tondeur_1984}, SkP \cite{Dobaczewski_1984}, SkI3, SkI4, and SkI5
\cite{RF95}, SLy4, SLy5, and SLy6 \cite{Chabanat_1998}, SKX, SKXm, and
SKXce \cite{Brown_1998}, SkO and SkO$'$ \cite{Reinhard_1999}, MSk1 and
MSk3 \cite{Tondeur_2000}, MSk9 \cite{Goriely_2001}, SV-bas, SV-K218,
SV-kap00, SV-mas07, SV-sym34, and SV-min \cite{Kluepfel_2009},
SV-m56k6 and SV-m64k6 \cite{Lyutorovich_2012}, and SAMi
\cite{Roca-Maza_2012}.

Here it should be noted that the time-odd densities and currents are equal to zero
in the ground states of the even-even nuclei \cite{Engel_1975}.
So, the constants $\,C_T^{s}$, $\,C_T^{s,\alpha}$, and $\,C_T^{\Delta s}$
do not affect the ground-state properties of these nuclei
and the mean field deduced by making use of Eq.~(\ref{sccond}).
Nevertheless, these constants can have an impact on the characteristics of
the excited states of the even-even nuclei because in the general case the
respective terms of the functional $E_\mathrm{int}$
give the nonzero contribution to the residual interaction according to
Eqs. (\ref{sccond}), (\ref{ESk3}), (\ref{Eint}), and (\ref{Eden}),
even if the time-odd densities and currents are equal to zero.
This circumstance allows us to change the constants
$\,C_T^{s}$, $\,C_T^{s,\alpha}$, and $\,C_T^{\Delta s}$
(assuming that they are the independent parameters)
for the purpose of description of nuclear excitations without affecting
the ground state and the self-consistent mean field.

It is known that the parameters
$\,C_T^{s}$, $\,C_T^{s,\alpha}$, and $\,C_T^{\Delta s}$ in most cases have
little influence on the characteristics of the natural parity excitations,
but in some cases can lead to the spin instability in the self-consistent RPA and
extended RPA calculations.
In particular for this reason sometimes (including our recent papers
\cite{Lyutorovich_2015,Lyutorovich_2016,Tselyaev_2016,Tselyaev_2017,Tselyaev_2018})
they are set to be equal to zero, while the other $C$-constants
are determined by the Skyrme-force parameters according to Eqs.~(\ref{def:CT}).
However, this choice is not suitable for the self-consistent description
of the magnetic excitations which are the subject of the present paper.
In this case the terms of the functional $E_\mathrm{int}$ containing
the constants $\,C_T^{s}$, $\,C_T^{s,\alpha}$, and $\,C_T^{\Delta s}$
become relevant. In particular, from
Eqs. (\ref{sccond}), (\ref{ESk3}), (\ref{Eint}), and (\ref{Eden})
it follows that the terms containing $\,C_T^{s}$ yield
the term $V^s$ of the residual interaction $V$ having the form of the
Landau-Migdal ansatz
\be
V^s = C^{\vphu}_{\mbsu{N}} \bigl(\,g\;\bfsigma\cdot\bfsigmap
+ g'\,\bfsigma\cdot\bfsigmap\;\bftau\cdot\bftaup\,\bigr)
\label{vsmigdal}
\ee
where
\be
C^{\vphu}_{\mbsu{N}}\,g = 2\,C_0^{s}\,,\qquad
C^{\vphu}_{\mbsu{N}}\,g' = 2\,C_1^{s}\,,
\label{def:CMg}
\ee
$C^{\vphu}_{\mbsu{N}}$ is a normalization constant.
Just the parameters $g$ and $g'$ in Eq.~(\ref{vsmigdal})
are responsible for the description of the unnatural parity excitations
in the TFFS (see \cite{Migdal_1967,Ring_1973PLB,Borzov_1984}).
The method of determining the $C$-constants of the functional $E_\mathrm{int}$
adopted in the present paper is described in Sec.~\ref{sec:modpar}.

\subsection{The \boldmath M1 operator}
\label{sec:QM1}

The field operator $Q$ in the case of the M1 excitations has the following
(vector) form
\bea
\bfbq &=& \mu^{\vphu}_N\,\sqrt{\frac{3}{16\pi}}\,
\Bigl\{ (\gamma_n + \gamma_p\,)\,\bfsigma + \bfl
\nonumber\\
&+& \bigl[\,(1-2\xi_s)\,(\gamma_n - \gamma_p\,)\,\bfsigma
- (1-2\xi_{\,l})\,\bfl\,\bigr]\,\tau_3 \Bigr\}
\label{def:qnp}
\eea
where $\bfl$ is the single-particle operator of the angular momentum,
$\bfsigma$ and $\tau_3$ are the spin and isospin Pauli matrices,
respectively (with positive eigenvalue of $\tau_3$ for the neutrons),
$\mu^{\vphu}_N = e\hbar/2m_p c$ is the nuclear magneton, $\gamma_p =
2.793$ and $\gamma_n = -1.913$ are the spin gyromagnetic ratios,
$\xi_s$ and $\xi_{\,l}$ are the renormalization constants introduced
to simulate quenching of the M1 strength that is usually necessary
for the description of the experimental data.  The nonzero $\xi_s$ and
$\xi_{\,l}$ correspond to the effective operator $\bfbq$.  Their
standard values are (see \cite{Borzov_1984,Kamerdzhiev_2004})
\be
\xi_s = 0.1, \qquad  \xi_{\,l} = -0.03.
\label{xieff}
\ee
Zero values
\be
\xi_s = 0, \qquad  \xi_{\,l} = 0
\label{xizer}
\ee
correspond to the bare operator $\bfbq^{(0)}$.

Eq. (\ref{def:qnp}) can be represented as the result of the action
of the effective charge operator $e_q$ introduced in the TFFS \cite{Migdal_1967}
on the bare operator $\bfbq^{(0)}$, that is
\be
\bfbq = e_q \bfbq^{(0)}\,,
\label{def:qb}
\ee
where
\be
e_q = 1 - \ffrac{1}{2}\,(\,\xi_{\,l}\,\sigma^{\vphu}_0\,\sigma'_0 +
\xi_s\,\bfsigma\cdot\bfsigmap\,)\,\bftau\cdot\bftaup,
\label{def:eq}
\ee
and $\sigma^{\vphu}_0$ is the identity spin matrix.
According to the TFFS, the operator $e_q$ is universal, i.e. it should act
on all the external field operators $Q$ including the operators of the
electric type $Q^{\mbsu{e}}$ which are proportional to $\sigma^{\vphu}_0$.
From this it follows that if we impose the condition of the invariance
\be
e_q Q^{\mbsu{e}} = Q^{\mbsu{e}},
\label{eqqe}
\ee
we should set $\xi_{\,l}=0$. The actual values of this constant used in the
calculations of the magnetic excitations are very small and thus violate
the condition (\ref{eqqe}) only slightly.

\subsection{Numerical details}
\label{sec:numdet}

The equations of the RPA for the M1 excitations in $^{208}$Pb
were solved within the fully self-consistent scheme as described in Refs.
\cite{Lyutorovich_2015,Lyutorovich_2016,Tselyaev_2016}.

The single-particle basis was discretized by imposing the box boundary condition
with the box radius equal to 18~fm. The particle energies $\ve^{\vphu}_{p}$ were limited
by the maximum value $\ve^{\mbss{max}}_{p} = 100$ MeV. These conditions ensure
fulfillment of the RPA energy-weighted sum rule for the isoscalar $EL$ excitations
in $^{208}$Pb within 0.1~\% for $L \leqslant 8$.

\section{\boldmath M1 excitations in $^{208}\mbox{Pb}$ in RPA}
\label{sec:result}

\subsection{Defining the problem and observables}
\label{sec:problem}

\begin{figure}[h!]
\begin{center}
\includegraphics*[trim=0cm 0cm 0cm 2cm,clip=true,scale=0.4,angle=90]{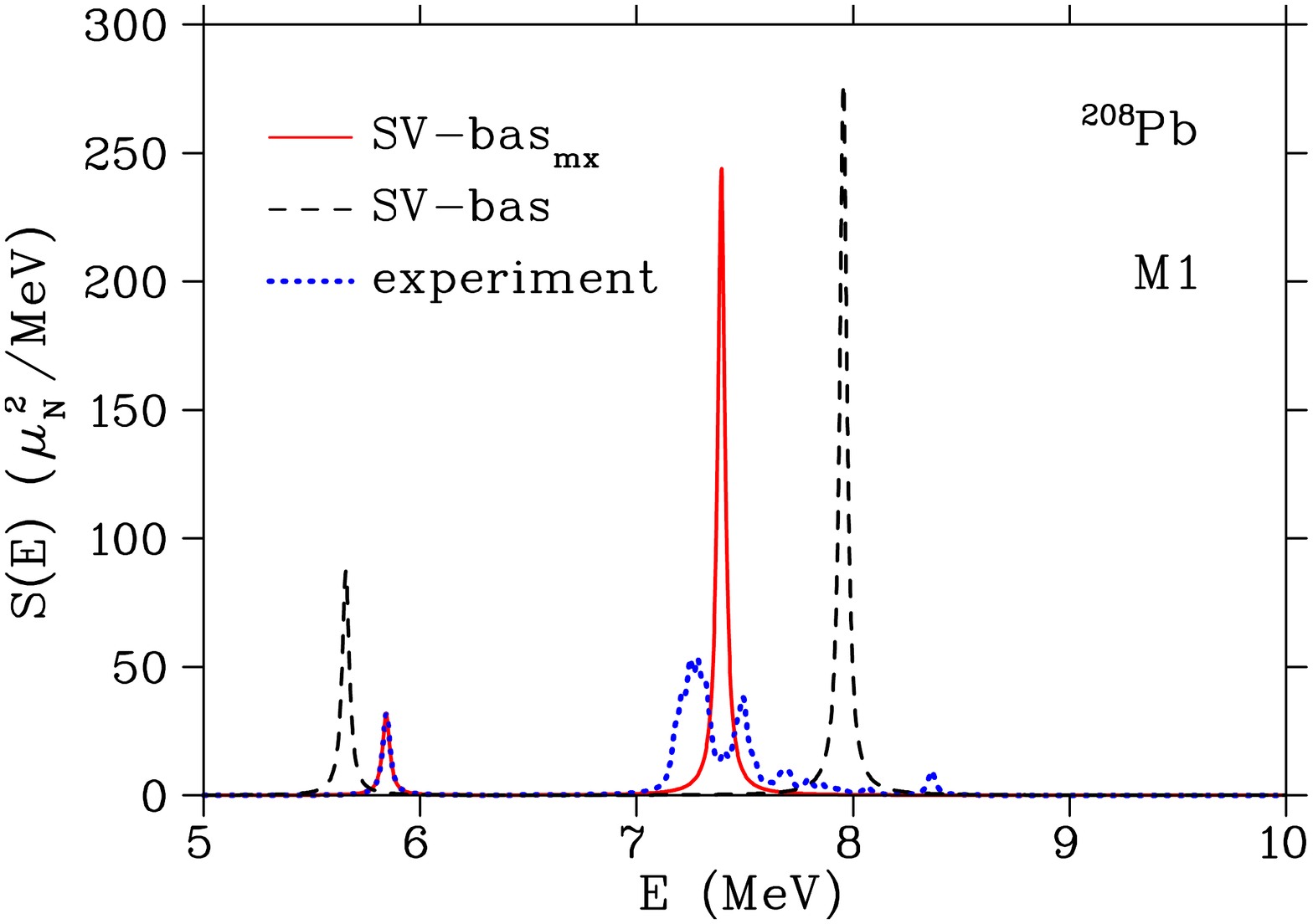}
\end{center}
\caption{\label{fig:m1rpa} Strength functions of the M1 excitations in
  $^{208}$Pb calculated within RPA using the parametrization SV-bas
  \cite{Kluepfel_2009} (black dashed line) and SV-bas$_{\mbss{mx}}$ as
  a modified variant thereof (red solid line) introduced in section
  \ref{sec:modpar}.  Experimental data taken from
  Refs.~\cite{Koehler_1987,Shizuma_2008} are shown by the blue dotted
  line. The low-lying M1 state is at 5.84 MeV, hidden below the result
  from SV-bas$_{\mbss{mx}}$.  The discrete peaks from RPA and the
  lower M1 mode have been broadened with a smearing parameter $\Delta
  = 20$ keV to represent a smooth distribution. }
\end{figure}

In order to illustrate the observables for the following survey, we
start with showing in Fig.~\ref{fig:m1rpa} the distribution of M1
strength in $^{208}$Pb calculated within self-consistent RPA based
on the Skyrme EDF with two different parametrizations and comparing it
with experimental data.  We employ here the discrete version of the
RPA because the single-particle continuum plays a minor role in
the considered case.
%
The strength functions were obtained by folding the discrete RPA
spectrum and the discrete experimental mode (lower M1 mode) with a
Lorentzian of half-width $\Delta=20$ keV.
The experimental data demonstrate the basic features of M1 strength in
$^{208}$Pb: there is a very narrow peak at lower energy $E_1=5.84$ MeV
and a broad resonance at $E_2=7.39$ MeV. The height of the lower peak
is characterized by its integrated $B_1(M1)$ strength. Experimental
mean energy and strength of the upper M1 resonance are computed from
moments $m_k=\Sigma_\nu\,B_\nu(M1)\,E_\nu^k$ summed/integrated in
the interval 6.6--8.1 MeV
with the probabilities $B_\nu(M1)$ and the
excitation energies $E_\nu$ taken from
Refs. \cite{Koehler_1987,Shizuma_2008}.
We indicate this procedure by the notation $\sum B(M1)$ for that
value.  Note that we do not include in this interval the state with
$E$ = 7.335 MeV (and possible $B(M1)$ = 1.8 $\mu^2_N$) from
Ref.~\cite{Shizuma_2008} because of the uncertainty with the
identification of its spin.  We also note that the chosen smearing
parameter $\Delta=20$ keV is sufficiently large to average out the fine
structure of the experimental spectrum which is not essential for our
analysis, but remains sufficiently small to resolve the spreading
widths.
The experimental strength distribution is composed from two data
  sets, below the neutron separation energy 7.37~MeV from
  \cite{Shizuma_2008} and above from \cite{Koehler_1987}.  It is thus
  not clear whether the dip between the peaks at 7.26 MeV and 7.47~MeV
  is a real effect. Inelastic proton scattering data
  \cite{Poltoratska12,Birkhan_2016} seems to indicate that the dip does not exist.
  Anyway, such detailed fragmentation structure cannot be described
  within RPA. Thus we use for comparison with RPA the average peak
  properties as explained above.  Altogether, we have four
observables $E_1$, $E_2$, $B_1(M1)$, and $\sum B(M1)$ which we use
henceforth to characterize the M1 modes in $^{208}$Pb.

Fig.~\ref{fig:m1rpa} shows theoretical results from two different
parametrizations. The parametrization SV-bas$_{\mbss{mx}}$ (which is
tuned to data such that theoretical and experimental curve for the
lower peak at 5.84 MeV coincide) stands at the end of our
investigations and will be discussed later.  The results for SV-bas
(computed here with the all spin-spin terms included,
i.e. $\eta^{\vphd}_{\Delta s}=1$) are typical for most of the
available Skyrme parametrizations. They agree qualitatively in that
theory also produces two dominant peaks in the correct energy range.
But the position of the peaks and their strengths differs too much
from the data. Reasons for that and possible cures will be discussed
in the following.

\subsection{State of the art}
\label{sec:mdt}

It is well known that the properties of the low-energy M1
excitations in $^{208}$Pb in the RPA are mainly determined by two
$ph$ configurations formed by the neutron's ($\nu$)
and proton's ($\pi$) spin-orbit doublets $1i_{11/2}-1i_{13/2}$ and
$1h_{9/2}-1h_{11/2}$. The main characteristics of these configurations
are the $ph$ energy differences. Since the single-particle spectra
produced by the various parametrizations of the Skyrme EDF are very
different one can trace correlations between the values of these
energy differences, parameters of the EDF, and the RPA results for the
M1 excitations in $^{208}$Pb.

\begin{figure}[h!]
\centerline{\includegraphics[width=\linewidth]{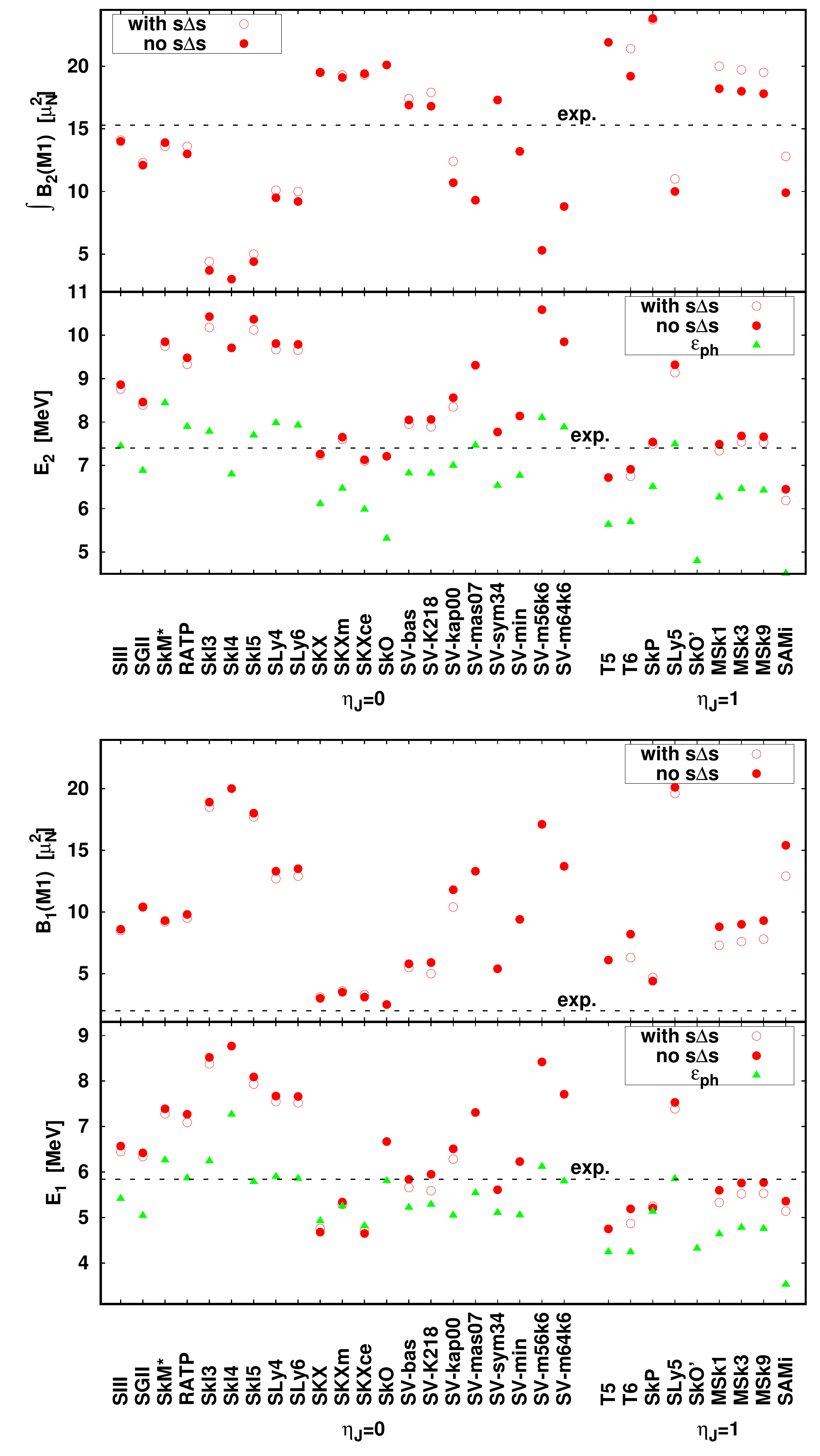}}
\caption{\label{fig:M1-traditional} RPA results for energies $E_n$ and
  $B_n(M1)$ values of the two leading M1-modes in $^{208}$Pb for a variety
  of published Skyrme parametrizations as listed at the end of section
  \ref{sec:EDF}. For the energies, we show also the leading $1ph$
  excitations $\ve^{\pi}_{ph}$ or $\ve^{\nu}_{ph}$, respectively.
  Experimental values are indicated by horizontal dotted lines. The
  parametrizations are grouped in those which omit tensor spin-orbit
  ($\eta^{\mbox{}}_J=0$, $C^J_T=0$) and those which use it
  ($\eta^{\mbox{}}_J=1$, $C^J_T\neq 0$).  RPA results are considered
  for two options concerning the spin gradient terms $\propto s\Delta
  s$: $\eta^{\mbox{}}_{\Delta s} = 1$ ($C_T^{\Delta s} \ne 0$) and
  $\eta^{\mbox{}}_{\Delta s} = 0$ ($C_T^{\Delta s}=0$).  }
\end{figure}

Let us introduce the notations
\bea
\ve^{\nu}_{ph} &=&
\ve^{\nu}_p(1i^{\vphu}_{11/2}) - \ve^{\nu}_h(1i^{\vphu}_{13/2})\,,
\label{def:phn}\\
\ve^{\pi}_{ph} &=&
\ve^{\pi}_p(1h^{\vphu}_{9/2}) -
\ve^{\pi}_h(1h^{\vphu}_{11/2})\,,
\label{def:php}
\eea
\be
\bar{\ve}^{\vphd}_{ph} = \frac{1}{2}\,
\bigl(\,\ve^{\nu}_{ph} + \ve^{\pi}_{ph}\,\bigr)\,,\qquad
\Delta \ve^{\vphd}_{ph} = \ve^{\nu}_{ph} - \ve^{\pi}_{ph}\,.
\label{def:dph}
\ee
The values of $\ve^{\nu}_{ph}$ and $\ve^{\pi}_{ph}$ along with the
energies and the reduced probabilities of the excitation of the
(isoscalar) $1^+_1$ state and the mean energies and the summed
strengths of the (isovector) M1 resonance in $^{208}$Pb calculated
within the self-consistent RPA for the parametrizations of the Skyrme
EDF indicated in Sec.~\ref{sec:EDF} are presented in
Figure~\ref{fig:M1-traditional}.
The effective M1 operator (\ref{def:qnp}) with the renormalization
constants $\xi_s$ and $\xi_{\,l}$ from Eq.~(\ref{xieff}) is used.
The shifts from mere $\ve_{ph}$ to the
  corresponding RPA energies $E_n$ indicate the strength of residual
  interaction in the M1 channel. It is generally smaller than for the
  giant resonances.  The figure reveals three main problems: First,
some Skyrme-EDF parametrizations used with all spin terms
[that means $\eta^{\mbox{}}_{\Delta s} = 1$ in Eqs. (\ref{def:CT})
and is denoted by open circles and the label ``with $\bfs\Delta\bfs$''
in Fig.~\ref{fig:M1-traditional}]
lead to spin instability (imaginary RPA solutions)
and thus have no entry in the plot (missing open circles).
Second,
the reduced probability $B_1(M1)$ of excitation of the first $1^+$
state significantly exceeds its experimental value for the most
parametrizations,
despite the quenching produced by the effective M1 operator.
Third, the mismatch starts already at the level of
pure $1ph$ energies $\ve^{\nu}_{ph}$ which are definitely too large
(upper panel) which can be tracked down to the fact that all
parametrizations give too large values of $\Delta \ve_{ph}$ as
compared to the experiment (see Figure~\ref{fig:spe-trends}).  As a
result, none of the parametrizations listed in
Figure~\ref{fig:M1-traditional} gives a satisfactory description of
both M1-modes simultaneously.  These problems were already found in earlier
publications and the spin-orbit coupling was identified as one
mechanism driving the M1 properties \cite{Nesterenko_2010}. We will
now discuss that in more detail and explore ways for a solution.

\subsection{Spin stability}
\label{sec:stab}

Spin stability is a crucial issue in the construction of Skyrme
parametrizations \cite{Str76a,Chabanat_1998}. The first is to check
the stability of bulk matter which is done easily in terms of the
LM parameters of the residual interaction.  The LM
parameters are related with the $C$-constants of the Skyrme-EDF by the
following equations (see, e.g., Refs.~\cite{Bender_2002,Chamel_2009})
\begin{subequations}
\label{def:LMP}
\bea
F^{\vphu}_0 &=& 2 N_0\,[\,C^{\rho}_0 +
\half (\alpha + 1)(\alpha + 2)C^{\rho,\alpha}_0 \rho^{\alpha}_{\mbsu{eq}} +
C^{\tau}_0 k^2_{\mbsu{F}}]\,,\qquad\;\;
\label{def:f0}\\
F'_0 &=& 2 N_0\,[\,C^{\rho}_1 +
C^{\rho,\alpha}_1 \rho^{\alpha}_{\mbsu{eq}} +
C^{\tau}_1 k^2_{\mbsu{F}}\,]\,,
\label{def:f0p}\\
G^{\vphu}_0 &=& 2 N_0\,[\,C^{s}_0 +
C^{s,\alpha}_0 \rho^{\alpha}_{\mbsu{eq}} -
C^{J}_0 k^2_{\mbsu{F}}\,]\,,
\label{def:g0}\\
G'_0 &=& 2 N_0\,[\,C^{s}_1 +
C^{s,\alpha}_1 \rho^{\alpha}_{\mbsu{eq}} -
C^{J}_1 k^2_{\mbsu{F}}\,]\,,
\label{def:g0p}\\
F^{\vphu}_1 &=& -2 N_0 C^{\tau}_0 k^2_{\mbsu{F}}\,,\;\,\quad
F'_1 = -2 N_0 C^{\tau}_1 k^2_{\mbsu{F}}\,,
\label{def:f1}\\
G^{\vphu}_1 &=& 2 N_0 C^{J}_0 k^2_{\mbsu{F}}\,,\qquad
G'_1 = 2 N_0 C^{J}_1 k^2_{\mbsu{F}}\,,
\label{def:g1}
\eea
\end{subequations}
where $N_0 = 2m^* k_{\mbsu{F}}/(\pi \hbar)^2$, $k_{\mbsu{F}} = (3\pi^2
\rho_{\mbsu{eq}} /2)^{1/3}$ is the Fermi momentum, and
$\rho_{\mbsu{eq}}$ is the equilibrium density of the infinite nuclear
matter (INM).  Eqs. (\ref{def:LMP}) coincide with the definitions of
Ref.~\cite{vanGiai_1981} if the $C$-constants are expressed through
the parameters of the Skyrme force by the standard formulas. However,
Eqs. (\ref{def:LMP}) produce $G_L$ and $G'_L$ at variance with
Ref.~\cite{vanGiai_1981} for those parametrizations in which the $J^2$
terms are omitted ($\eta^{\vphu}_J=0$ and $C^J_T = 0$) as noted in
\cite{Lesinski_2007}.  In particular, the parameters $G_1$ and $G'_1$
are exactly equal to zero if the $J^2$ terms are absent in the Skyrme
EDF.
To ensure stability, the LM parameters should satisfy the following
inequalities (see \cite{Migdal_1967})
\begin{subequations}
\label{LMPst}
\bea
&&\frac{F^{\vphu}_L}{2L+1} > -1\,,\qquad \frac{F'_L}{2L+1} > -1\,,
\label{LMPfst}\\
&&\frac{G^{\vphu}_L}{2L+1} > -1\,,\qquad \frac{G'_L}{2L+1} > -1 \,.
\label{LMPgst}
\eea
\end{subequations}

\begin{table*}[ht!]
\caption{\label{tab:LMP} Landau-Migdal parameters of the Skyrme-EDFs
  listed in Figure \ref{fig:M1-traditional}.  }
\begin{ruledtabular}
\begin{tabular}{lccccccccccccc}
 EDF & $\eta^{\vphu}_J$ & $x_W$ & $F^{\vphu}_0$ & $F'_0$ & $G^{\vphu}_0$ & $G'_0$
 & $F^{\vphu}_1$ & $F'_1$ & $G^{\vphu}_1$ & $G'_1$
 & $N^{-1}_0$ & $m^*/m$ & $k_{\mbsu{F}}$ \\
 &&&&&&&&&&& (MeV$\cdot$fm$^3$) && (fm$^{-1}$) \\
\hline
 SIII     & 0 & 1
 & $\hfm$0.31 & $\hfm$0.87 & $\hfm$0.54 & $\hfm$0.95 &    $-$0.71 & $\hfm$0.49
 &  0 &  0 & 207.8 & 0.76 & 1.29 \\
 SGII     & 0 & 1
 &    $-$0.23 & $\hfm$0.73 & $\hfm$0.62 & $\hfm$0.93 &    $-$0.64 & $\hfm$0.52
 &  0 &  0 & 196.1 & 0.79 & 1.33 \\
 SkM$^*$  & 0 & 1
 &    $-$0.23 & $\hfm$0.93 & $\hfm$0.33 & $\hfm$0.94 &    $-$0.63 & $\hfm$0.62
 &  0 &  0 & 194.6 & 0.79 & 1.33 \\
 RATP     & 0 & 1
 &    $-$0.28 & $\hfm$0.59 & $\hfm$0.63 & $\hfm$0.89 &    $-$1.00 & $\hfm$0.56
 &  0 &  0 & 230.2 & 0.67 & 1.33 \\
 T5       & 1 & 1
 &    $-$0.10 & $\hfm$1.96 &    $-$0.88 & $\hfm$0.05 &    $-$0.00 &    $-$0.00
 &  0.97 &  0.97 & 152.3 & 1.00 & 1.34 \\
 T6       & 1 & 1
 & $\hfm$0.06 & $\hfm$1.43 &    $-$0.22 & $\hfm$0.18 &    $-$0.00 &    $-$0.00
 &  0.86 &  0.86 & 153.3 & 1.00 & 1.34 \\
 SkP      & 1 & 1
 &    $-$0.10 & $\hfm$1.42 &    $-$0.23 & $\hfm$0.06 & $\hfm$0.00 & $\hfm$1.05
 & $-$0.18 &  0.97 & 152.7 & 1.00 & 1.34 \\
 SkI3     & 0 & 0
 &    $-$0.32 & $\hfm$0.65 & $\hfm$1.90 & $\hfm$0.85 &    $-$1.27 &    $-$0.84
 &  0 &  0 & 267.2 & 0.58 & 1.33 \\
 SkI4     & 0 & $\!\!\!\!-$0.99
 &    $-$0.27 & $\hfm$0.56 & $\hfm$1.77 & $\hfm$0.88 &    $-$1.05 &    $-$0.57
 &  0 &  0 & 236.4 & 0.65 & 1.33 \\
 SkI5     & 0 & 1
 &    $-$0.32 & $\hfm$0.76 & $\hfm$1.79 & $\hfm$0.85 &    $-$1.26 &    $-$0.84
 &  0 &  0 & 267.7 & 0.58 & 1.32 \\
 SLy4     & 0 & 1
 &    $-$0.28 & $\hfm$0.81 & $\hfm$1.39 & $\hfm$0.90 &    $-$0.92 &    $-$0.40
 &  0 &  0 & 221.2 & 0.69 & 1.33 \\
 SLy5     & 1 & 1
 &    $-$0.28 & $\hfm$0.81 & $\hfm$1.12 &    $-$0.14 &    $-$0.91 &    $-$0.39
 &  0.25 &  1.04 & 220.1 & 0.70 & 1.33 \\
 SLy6     & 0 & 1
 &    $-$0.28 & $\hfm$0.80 & $\hfm$1.41 & $\hfm$0.90 &    $-$0.93 &    $-$0.41
 &  0 &  0 & 223.0 & 0.69 & 1.33 \\
 SKX      & 0 & 0
 & $\hfm$0.24 & $\hfm$1.56 &    $-$0.46 & $\hfm$1.04 &    $-$0.02 & $\hfm$0.98
 &  0 &  0 & 156.1 & 0.99 & 1.32 \\
 SKXm     & 0 & 0
 & $\hfm$0.05 & $\hfm$1.47 &    $-$0.29 & $\hfm$1.02 &    $-$0.10 & $\hfm$0.87
 &  0 &  0 & 159.4 & 0.97 & 1.33 \\
 SKXce    & 0 & 0
 & $\hfm$0.24 & $\hfm$1.52 &    $-$0.45 & $\hfm$1.04 & $\hfm$0.02 & $\hfm$1.01
 &  0 &  0 & 154.1 & 1.01 & 1.32 \\
 SkO      & 0 & $\!\!\!\!-$1.13
 &    $-$0.10 & $\hfm$1.33 & $\hfm$0.48 & $\hfm$0.98 &    $-$0.31 & $\hfm$0.16
 &  0 &  0 & 171.2 & 0.90 & 1.33 \\
 SkO$'$   & 1 & $\!\!\!\!-$0.58
 &    $-$0.10 & $\hfm$1.33 &    $-$1.61 & $\hfm$0.79 &    $-$0.31 & $\hfm$0.09
 &  2.16 &  0.19 & 171.3 & 0.90 & 1.33 \\
 MSk1     & 1 & 1
 & $\hfm$0.07 & $\hfm$1.47 &    $-$0.18 & $\hfm$0.25 &    $-$0.00 &    $-$0.00
 &  0.78 &  0.78 & 154.3 & 1.00 & 1.33 \\
 MSk3     & 1 & 1
 & $\hfm$0.07 & $\hfm$1.30 &    $-$0.00 & $\hfm$0.27 &    $-$0.00 &    $-$0.00
 &  0.77 &  0.77 & 154.3 & 1.00 & 1.33 \\
 MSk9     & 1 & 1
 & $\hfm$0.07 & $\hfm$1.30 &    $-$0.02 & $\hfm$0.25 &    $-$0.00 &    $-$0.00
 &  0.78 &  0.78 & 154.3 & 1.00 & 1.33 \\
 SV-bas   & 0 & 0.55
 &    $-$0.05 & $\hfm$1.20 & $\hfm$0.00 & $\hfm$0.99 &    $-$0.30 & $\hfm$0.78
 &  0 &  0 & 170.8 & 0.90 & 1.33 \\
 SV-K218  & 0 & 0.45
 &    $-$0.12 & $\hfm$1.18 & $\hfm$0.02 & $\hfm$0.99 &    $-$0.30 & $\hfm$0.77
 &  0 &  0 & 170.3 & 0.90 & 1.34 \\
 SV-kap00 & 0 & 1.33
 &    $-$0.05 & $\hfm$1.20 & $\hfm$1.08 & $\hfm$0.99 &    $-$0.30 &    $-$0.30
 &  0 &  0 & 170.8 & 0.90 & 1.33 \\
 SV-mas07 & 0 & 1.02
 &    $-$0.26 & $\hfm$0.71 & $\hfm$1.16 & $\hfm$0.90 &    $-$0.90 &    $-$0.06
 &  0 &  0 & 219.5 & 0.70 & 1.33 \\
 SV-sym34 & 0 & 0.29
 &    $-$0.04 & $\hfm$1.50 &    $-$0.29 & $\hfm$0.99 &    $-$0.30 & $\hfm$0.78
 &  0 &  0 & 170.9 & 0.90 & 1.33 \\
 SV-min   & 0 & 0.83
 &    $-$0.05 & $\hfm$1.37 & $\hfm$0.58 & $\hfm$1.01 &    $-$0.14 & $\hfm$0.07
 &  0 &  0 & 160.9 & 0.95 & 1.34 \\
 SV-m56k6 & 0 & 0.79
 &    $-$0.35 & $\hfm$0.24 & $\hfm$1.78 & $\hfm$0.84 &    $-$1.33 &    $-$0.33
 &  0 &  0 & 277.4 & 0.56 & 1.33 \\
 SV-m64k6 & 0 & 1.10
 &    $-$0.30 & $\hfm$0.40 & $\hfm$1.30 & $\hfm$0.87 &    $-$1.09 & $\hfm$0.05
 &  0 &  0 & 242.3 & 0.64 & 1.33 \\
 SAMi     & 1 & 0.31
 &    $-$0.25 & $\hfm$0.56 & $\hfm$0.15 & $\hfm$0.35 &    $-$0.97 & $\hfm$0.05
 &  1.03 &  0.54 & 228.0 & 0.68 & 1.33 \\
\end{tabular}
\end{ruledtabular}
\end{table*}
Table~\ref{tab:LMP} shows the LM parameters corresponding to the
Skyrme-EDF parametrizations listed in Figure \ref{fig:M1-traditional}.
The values of the spin-orbit parameter $x_W$ which will be discussed
in Sec.~\ref{sec:xw} are also given.
The conditions (\ref{LMPst}) are fulfilled for all parameters from
Table~\ref{tab:LMP} except for the parameter $G^{\vphu}_0$ of SkO$'$.
However, as can be seen from Figure \ref{fig:M1-traditional}, the
parametrizations T5, SkI4, SkO, SV-mas07, SV-sym34, SV-min, SV-m56k6,
and SV-m64k6, for which the INM is stable, lead to the spin
instability of the ground state of $^{208}$Pb in the case of
$\eta^{\vphd}_{\Delta s} = 1$, in spite of bulk stability as proven by
Table~\ref{tab:LMP}.  This instability appears only in certain finite
nuclei and is generated by the spin surface terms
$\propto{C}_T^{\Delta s}$, not contained in Eqs.~(\ref{def:LMP}) for
the LM parameters (see also Ref. \cite{Pastore_2015} where this
question is discussed in more detail).  On the other hand, Figure
\ref{fig:M1-traditional} shows that the inclusion of the terms
proportional to $C_T^{\Delta s}$ into the Skyrme EDF usually decreases
the energy of the $1^+_1$ state (compare open with filled circles).
Exceptions from this general trend are SkP, SKX, and SKXce for which
$E(1^+_1)$ slightly increases if $\eta^{\vphd}_{\Delta s} = 1$.  If
the downshift by the $C_T^{\Delta s}$ terms grows too large, it drives
the finite nucleus to instability.  All the Skyrme-EDF
parametrizations shown in Figure \ref{fig:M1-traditional} except for
SkO$'$ provide a stable ground state for $^{208}$Pb in case of
$\eta^{\vphd}_{\Delta s} = 0$ which is in agreement with the INM
properties resulting from Table~\ref{tab:LMP}.

Note that the instability generated by the EDF SkO$'$ disappears in the modified
parametrization SkO$'_{\mbsu{m}}$,
in which the $C$-constants are determined by Eqs.~(\ref{def:CT})
with $\eta^{\vphd}_{s}$ = $\eta^{\vphd}_{s,\alpha}$ = $\eta^{\vphd}_{\Delta s}$ = 0,
$C^{\vphu}_{\mbsu{N}} = 300$ MeV$\cdot$fm$^3$, $g=0.891$, and $g'=1.39$.
In this case we have $G^{\vphu}_0 = -0.60$, $G'_0$ = 2.24.
The parameters $F^{\vphu}_{0,1}$, $F'_{0,1}$, $G^{\vphu}_1$, and $G'_1$ are not changed.
Thus, the nuclear matter becomes stable.
The parameters $g$ and $g'$ in SkO$'_{\mbsu{m}}$ have been adjusted to reproduce within
the RPA the experimental energies of the M1 excitations in $^{208}$Pb,
$E_1$ = 5.84 MeV and $E_2$ = 7.39 MeV.
The $B(M1)$ values for the $1^+_1$ state and the isovector
$M1$ resonance in $^{208}$Pb in this parametrization are equal to 1.9 $\mu^2_N$
and 16.9 $\mu^2_N$, respectively.

\subsection{The impact of spin-orbit parameters}
\label{sec:xw}

\begin{figure}[h!]
\centerline{\includegraphics[width=1.0\linewidth]{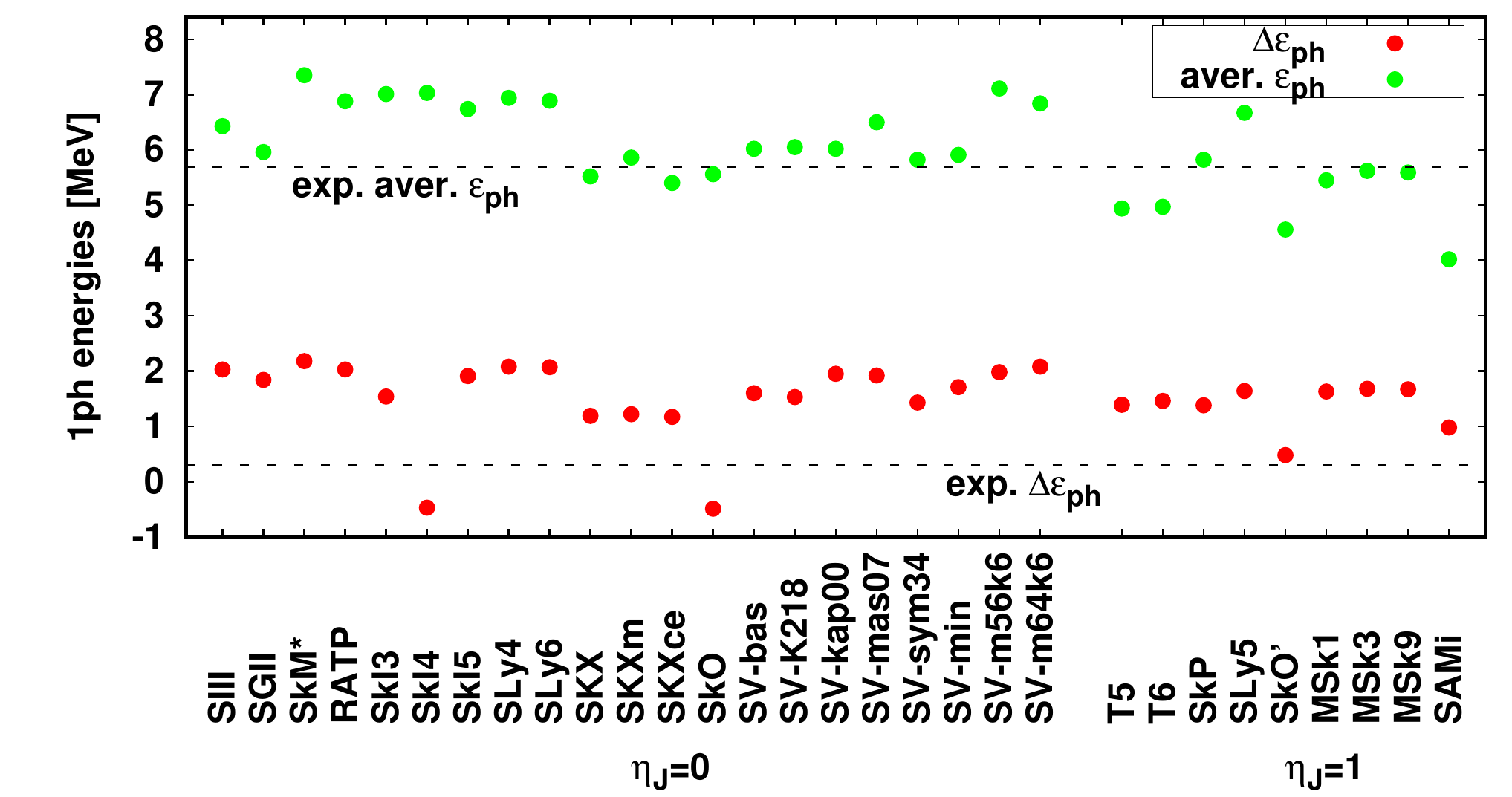}}
\caption{\label{fig:spe-trends} Average $1ph$ energy and difference as
  defined in Eq. (\ref{def:dph}) for the same selection of published
  Skyrme parametrizations as in Figure~\ref{fig:M1-traditional}.
  Experimental values are indicated by horizontal dotted lines. }
\end{figure}

Figure~\ref{fig:M1-traditional} indicates that problems appear already
at the level of the  $1ph$ energies. This becomes even more
apparent when looking at the average and difference $1ph$ energies
(\ref{def:dph}) as shown in Figure~\ref{fig:spe-trends}.  First,
$\Delta \ve_{ph}$ exceeds for most parametrizations the experimental
value (0.29~MeV) by a factor of 3.4 (SAMi) to 7.5 (SkM$^*$), except
for SkI4, SkO, and SkO$'$ for which the spin-orbit parameter is
$x_W < 0$ (see Table~\ref{tab:LMP}).
Second, for the
parametrizations with $x_W > 0$, the value of $B_1(M1)$ calculated
with $\eta^{\vphd}_{\Delta s} = 0$ is greater than its experimental
value (2.0 $\mu^2_N$) by a factor of 2.2 (SkP) to 10 (SLy5).
This together suggests that the values of $x_W$ and $\Delta
  \ve_{ph}$ are key agents determining the RPA results for the M1
  excitations in $^{208}$Pb.

\begin{figure}[h!]
\begin{center}
\includegraphics*[trim=2.5cm 0cm 0cm 0cm,clip=true,scale=0.5,angle=0]{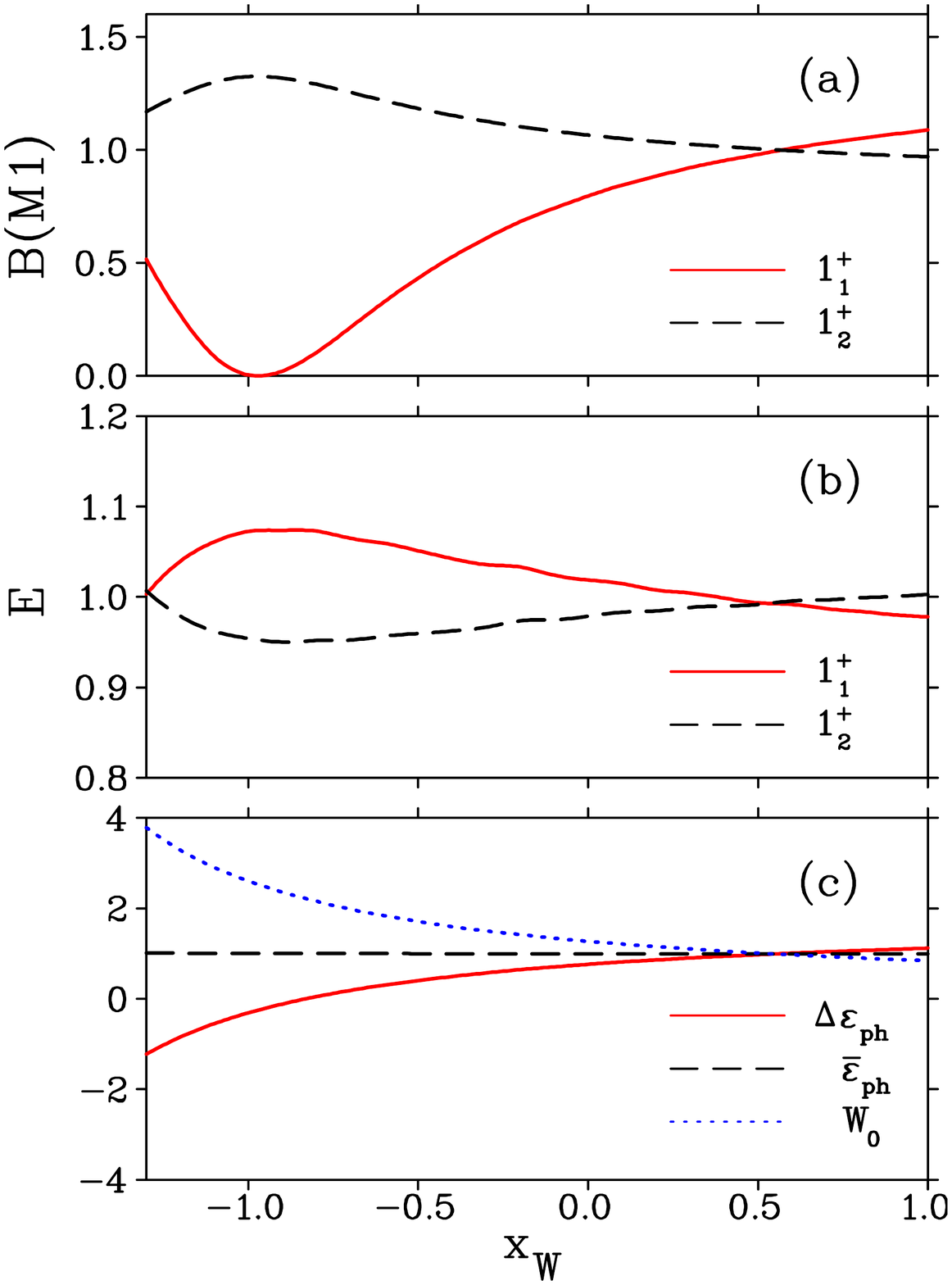}
\end{center}
\caption{\label{fig:xwf}
Dependence of the characteristics of M1 excitations in $^{208}$Pb on the parameter
$x_W$ of the Skyrme EDF. Parametrization SV-bas \cite{Kluepfel_2009} is used.
(a) The reduced probabilities $B(M1)$ of the excitation of the first (solid red line)
and second (dashed black line) $1^+$ states calculated in the RPA.
(b) Same as in panel (a) but for the energies of these states.
(c) The values of the energy differences $\Delta \ve_{ph}$ (solid red line) and
$\bar{\ve}_{ph}$ (dashed black line), Eqs.~(\ref{def:dph}), and the spin-orbit
parameter $W_0$ (dotted blue line).
All the quantities are given in units of their values obtained for the
original parametrization \cite{Kluepfel_2009}. See text for more details.
}
\end{figure}

To explore this further, we consider simultaneous variation of the
spin-orbit parameters $x_W$ and $W_0$. To that end, we start from the
set SV-bas \cite{Kluepfel_2009}, vary $x_W$, keeping all other model
parameters frozen, and tune $W_0$ to reproduce the SHF binding energy
of $^{208}$Pb at its experimental value 1636.43~MeV within the
accuracy of 0.2~MeV. This is done for the option
$\eta^{\vphd}_{\Delta s} = 1$.
Figure~\ref{fig:xwf} shows the dependence of the RPA results for the
first and second $1^+$ states in $^{208}$Pb on the parameter $x_W$
obtained in this way.    The respective values of $\Delta \ve_{ph}$,
$\bar{\ve}_{ph}$, and $W_0$ are also shown.
All these quantities are given in units of their values obtained for
the original set SV-bas \cite{Kluepfel_2009} and
shown in Figures~\ref{fig:M1-traditional} and \ref{fig:spe-trends}
[$B_1(M1)$ = 5.5 $\mu^2_N$, $B_2(M1)$ = 17.4 $\mu^2_N$,
$E_1$ = 5.66~MeV, $E_2$ = 7.95~MeV,
$\Delta \ve_{ph}$ = 1.60~MeV, $\bar{\ve}_{ph}$ = 6.02~MeV]
and the value $W_0$ = 124.634 MeV$\cdot$fm$^5$.
The  $B_1(M1)$ shows the strongest dependence on
$x_W$.  In fact, one can obtain any value of
$B_1(M1) <$ 6 $\mu^2_N$ by decreasing the parameter $x_W$.  The
experimental value $B_1(M1)$ = 2 $\mu^2_N$ is obtained at $x_W <$
0.  The values of $E_1$, $E_2$, and $B_2(M1)$ depend on
$x_W$ to much lesser extent.  The energy difference $\Delta \ve_{ph}$
also shows a strong dependence
on $x_W$, while the value of $\bar{\ve}_{ph}$ is
nearly constant (it is changed within 2.2\% in the considered interval
of $x_W$).  The trend
of  $\Delta \ve_{ph}$ with $x_W$ is
 monotonous. This allows to transform the dependencies shown in
panels (a) and (b) of Fig.~\ref{fig:xwf} into analogous dependencies
on $\Delta \ve_{ph}$.
\begin{figure}[h!]
\begin{center}
\includegraphics*[trim=0cm 0cm 0cm 2cm,clip=true,scale=0.35,angle=90]{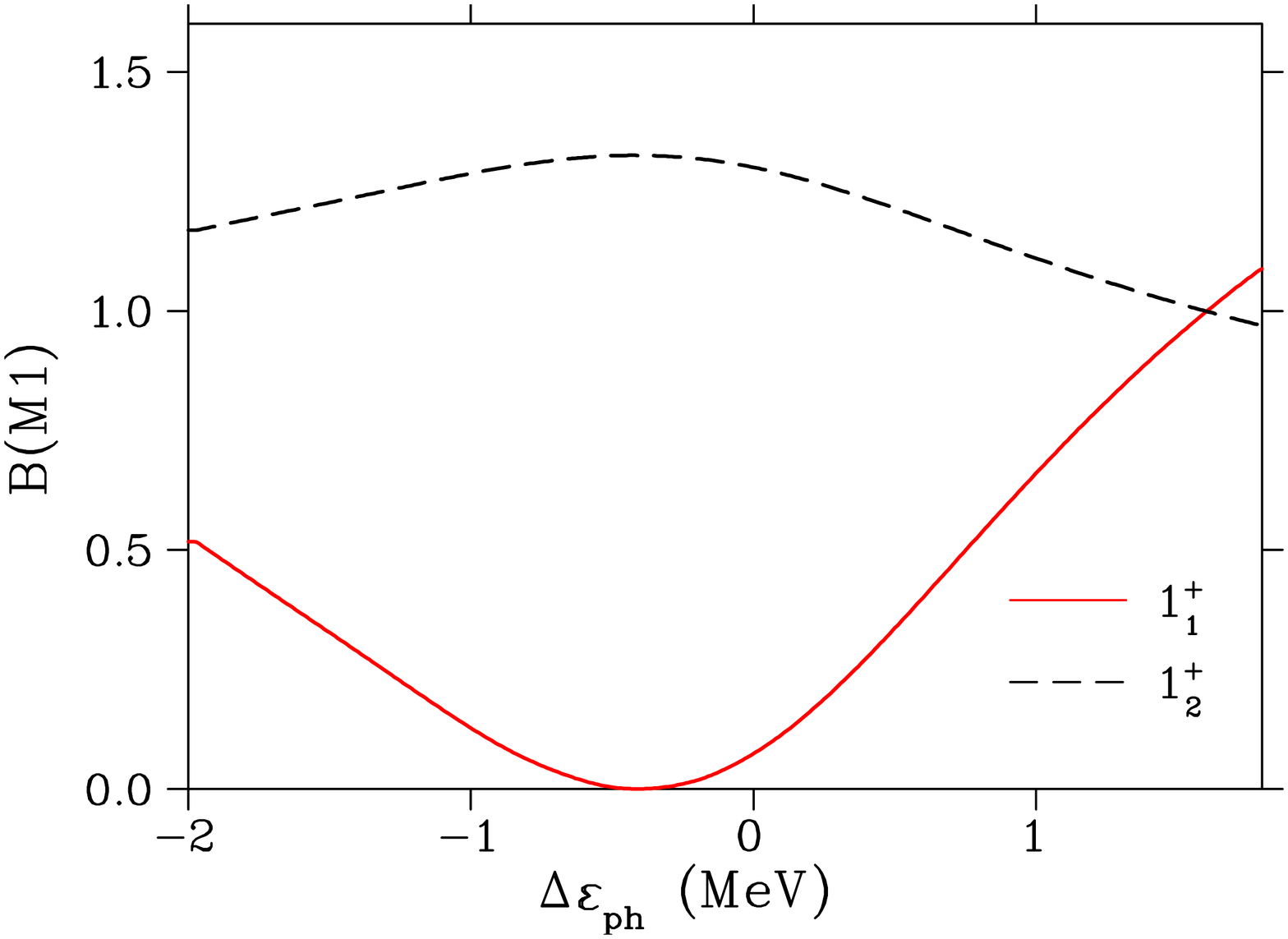}
\end{center}
\caption{\label{fig:def}
Same as in Fig. \ref{fig:xwf} but for the dependence of the reduced probabilities
$B(M1)$ on the value of the energy difference $\Delta \ve_{ph}$.
}
\end{figure}
The results are shown in Fig.~\ref{fig:def}, where again we see the
crucial dependence of $B_1(M1)$ on $\Delta \ve_{ph}$ at the
constant $\bar{\ve}_{ph}$.  This dependence explains why the
parametrization SkO$'_{\mbsu{m}}$ introduced in Sec.~\ref{sec:stab}
gives nice agreement with the experimental value of $B_1(M1)$: it
has negative $x_W = -0.58$ and thus a value of $\Delta \ve_{ph}$ = 0.48 MeV
which is closest to the experimental value 0.29 MeV.  The other Skyrme-EDF
parametrizations have generally too large $\Delta \ve_{ph}$ which
leads to significant overestimation of the $B_1(M1)$.  The impact
of the value of $\Delta \ve_{ph}$ on the properties of M1
excitations in $^{208}$Pb was pointed out in
\cite{Vesely_2009,Nesterenko_2010}.

\section{Toward better reproduction of M1 modes}
\label{sec:modpar}

The results presented in Sec.~\ref{sec:xw} show that spin-orbit
parameters are most decisive for the M1-modes. And, of course, the
parameters of the spin-spin terms play an equally important role.
This motivates us to check the chances to find a Skyrme functional in
standard form which provides a good description of M1-modes together
with traditionally good modeling of ground state properties. At
present stage, it is too early to launch a fully fledged least-squares
fitting scheme \cite{Kluepfel_2009,Kortelainen_2010,Dob14a}
particularly because a high precision RPA computation of M1-modes is
far too expensive.  Thus, for a first exploration, we employ a
simple-minded, restricted fitting procedure: We start from a given
Skyrme parametrization, keep all model parameters at their given value
except for the spin-orbit parameters $C_T^{\nabla J}$ (alias $x_W$,
$W_0$) and the spin-spin parameters $\,C_T^{s}$, $\,C_T^{s,\alpha}$,
and $\,C_T^{\Delta s}$. The spin-spin  parameters play no role for
ground states of even-even nuclei. Thus we exploit here the freedom
of not yet fixed parameters. However, the spin-orbit parameters
enter ground state properties. Here we have to check that re-tuning
does not destroy ground-state quality.

To keep the number of free spin-spin parameters low, we set
$\eta^{\vphd}_{s,\alpha}$ = $\eta^{\vphd}_{\Delta s}$ = 0 and
determine $\,C_T^{s}$ by Eqs.~(\ref{def:CT}) with $\eta^{\vphd}_{s}$ = 0
and the fitting parameters $g$ and $g'$ at $C^{\vphu}_{\mbsu{N}} =
300$ MeV$\cdot$fm$^3$.
After all, we have four free parameters $x_W$, $W_0$, $g$, and $g'$
which are determined by adjusting four observables in $^{208}$Pb: the
binding energy and the RPA results for the M1 energies $E_1$ and
$E_2$ and the transition probability $B_1(M1)$ to their
experimental values.
Note that here we use, as before,
the effective M1 operator (\ref{def:qnp}) with the renormalization
constants $\xi_s$ and $\xi_{\,l}$ from Eq.~(\ref{xieff}).
This fitting procedure is applied to a subset of
the parametrizations shown in Figure~\ref{fig:M1-traditional}.  The
modified parametrizations thus obtained are marked by an index ``m''.
\begin{figure*}
\centerline{\includegraphics[width=1.0\linewidth]{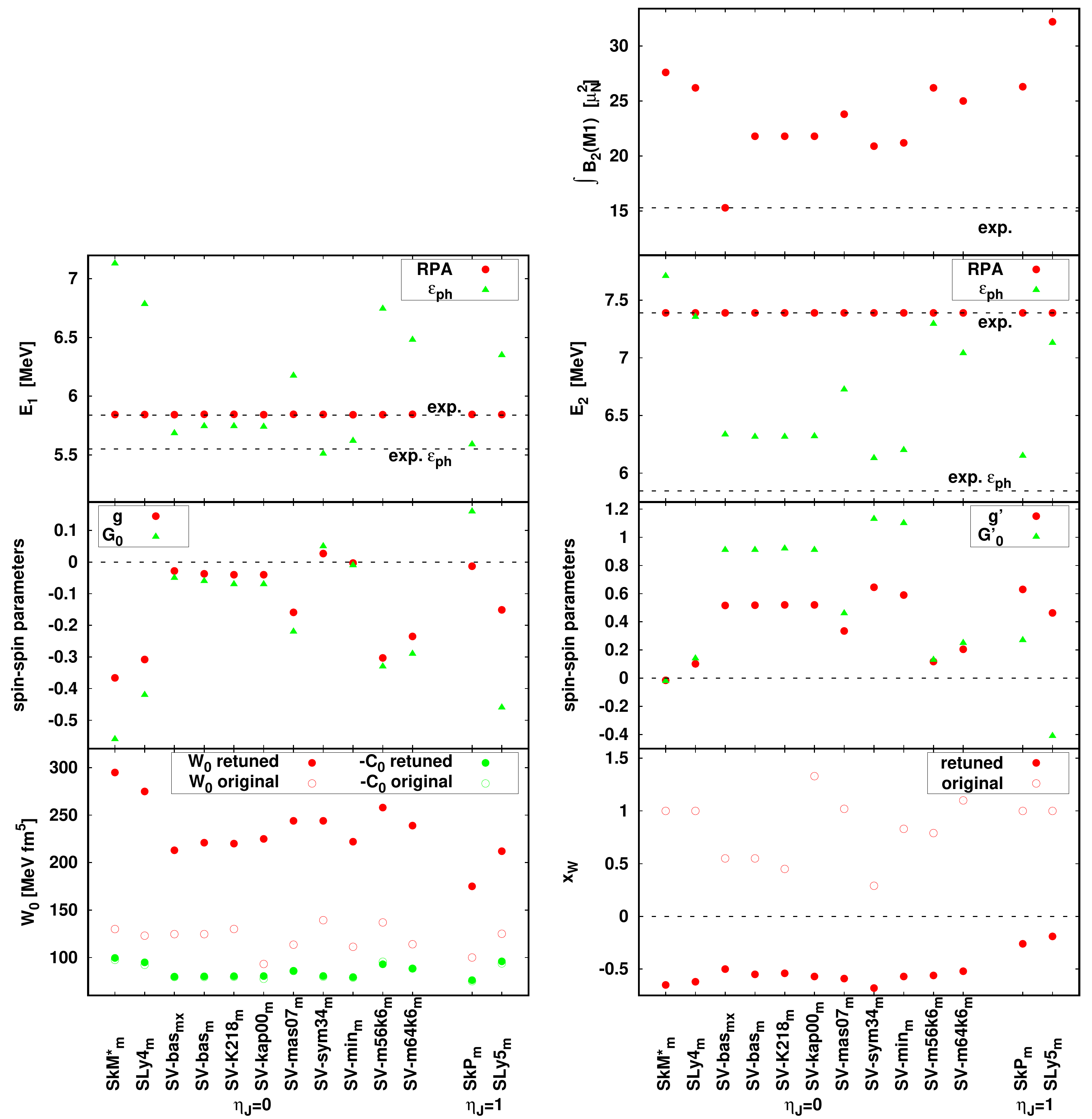}}
\caption{\label{fig:M1-refit} Results for the re-tuned
  parametrizations. Lower two panels: spin-orbit parameters $x_W$,
  $W_0$ (filled circles), and $-C_0^{\nabla J}$ (abbreviated $-C_0$ in
  the legend) together with their original values (open circles);
  spin-spin LM parameters $G_0$ and $G'_0$ together with the
  interaction parameters $g$ and $g'$ defined in Eqs. (\ref{def:CT}).
  Upper two panels: the M1 energies $E_1$ and $E_2$ together with
  their corresponding $1ph$ energies $\ve^{\pi}_{ph}$ and
  $\ve^{\nu}_{ph}$ and the $B(M1)$ strength for the upper M1 mode
  integrated over the interval 6.6--8.1 MeV. Experimental values are
  indicated by horizontal faint dashed lines. }
\end{figure*}
Resulting re-tuned model parameters and properties of M1-modes are
shown in Figure~\ref{fig:M1-refit}
and the corresponding re-tuned
spin-orbit and spin-spin parameters are given in quantitative detail
in Table \ref{tab:xw}.
\begin{table}
\caption{\label{tab:xw}
Parameters $\eta_J$, $x_W$, $W_0$, $g$, and $g'$ of the modified Skyrme EDFs.
Parameters $g$ and $g'$ of the Landau-Migdal interaction (\ref{vsmigdal})
are taken from Ref.~\cite{Migli_1991}.
}
\begin{ruledtabular}
\begin{tabular}{lccccc}
EDF & $\eta^{\vphu}_J$ & $x_W$ & $W_0$ & $g$ & $g'$ \\
&&& (MeV$\cdot$fm$^5$) &&\\
\hline
 SkM$^*_{\mbss{m}}$    & 0 & $-$0.65 & 295 & $-$0.366    & $-$0.015  \\
 SLy4$_{\mbss{m}}$     & 0 & $-$0.62 & 275 & $-$0.308    & $\hfm$0.102\\
 SV-bas$_{\mbss{mx}}$  & 0 & $-$0.50 & 213 & $-$0.028    & $\hfm$0.516\\
 SV-bas$_{\mbss{m}}$   & 0 & $-$0.55 & 221 & $-$0.037    & $\hfm$0.518\\
 SV-K218$_{\mbss{m}}$  & 0 & $-$0.54 & 220 & $-$0.040    & $\hfm$0.520\\
 SV-kap00$_{\mbss{m}}$ & 0 & $-$0.57 & 225 & $-$0.040    & $\hfm$0.520\\
 SV-mas07$_{\mbss{m}}$ & 0 & $-$0.59 & 244 & $-$0.159    & $\hfm$0.335\\
 SV-sym34$_{\mbss{m}}$ & 0 & $-$0.68 & 244 & $\hfm$0.027 & $\hfm$0.645\\
 SV-min$_{\mbss{m}}$   & 0 & $-$0.57 & 222 & $-$0.003    & $\hfm$0.590\\
 SV-m56k6$_{\mbss{m}}$ & 0 & $-$0.56 & 258 & $-$0.303    & $\hfm$0.118\\
 SV-m64k6$_{\mbss{m}}$ & 0 & $-$0.52 & 239 & $-$0.235    & $\hfm$0.205\\
 SkP$_{\mbss{m}}$      & 1 & $-$0.26 & 175 & $-$0.013    & $\hfm$0.630\\
 SLy5$_{\mbss{m}}$     & 1 & $-$0.19 & 212 & $-$0.151    & $\hfm$0.463\\
 Landau-Migdal &&&
 & $\hfm$0.1$\hfz\hfz$ & $\hfm$0.75$\hfz$\\
\end{tabular}
\end{ruledtabular}
\end{table}
As expected from the exploration in section \ref{sec:xw}, all re-tuned
$x_W$ parameters are negative, most of them in the interval between
$-0.6$ and $-0.5$.  Exceptions are SkP$_{\mbss{m}}$ and
SLy5$_{\mbss{m}}$ which have higher $x_W$ due to the $J^2$ terms in
these parametrizations which contribute also to the single-particle
spin-orbit potential. The re-tuned parameters $W_0$ are all rather
large. This seemingly happens to compensate the negative $x_W$. The
left lower panel
of Figure~\ref{fig:M1-refit}
shows also the isoscalar spin-orbit parameter
$C_0^{\nabla J}=-\ffrac{1}{4}(2+x_W)W_0$. This combination shows much
less variations over the different forces and, in particular, remains
practically unmodified by re-tuning. It is the isovector spin-orbit
term proportional to $C_1^{\nabla J}=-\ffrac{1}{4} x_W W_0$
which makes the difference. Seeing the dramatic differences in
spin-orbit parameters, one wonders what happens to the overall quality
of the parametrization. This question will be addressed farther below.

The spin-spin coupling parameters $g$ and $g'$ show some correlation
with the effective mass $m^*/m$ of a parametrization.  The sets
SkP$_{\mbss{m}}$, SV-bas$_{\mbss{m}}$, SV-K218$_{\mbss{m}}$,
SV-kap00$_{\mbss{m}}$, SV-sym34$_{\mbss{m}}$, and SV-min$_{\mbss{m}}$
all having $m^*/m\approx 1$ have similar values which are also close
to the values $g = 0.1$, $g' = 0.75$ used previously in the
non-self-consistent TFFS (see \cite{Migli_1991}) while the other
parametrizations having lower $m^*/m$ also produce lower $g$ and $g'$.

The LM parameters $G^{\vphu}_0$ and $G'_0$ in
Figure~\ref{fig:M1-refit} stay all safely above $-1$ and thus lead to
stable INM which also persists in finite nuclei because the modified
parametrizations set the critical gradient spin term to zero.

The RPA energies $E_n$ stay by construction at the experimental
values. We show them (second panels from above) to illustrate the span
toward the pure $1ph$ energies
$\ve^{\pi}_{ph}$ (left) and $\ve^{\nu}_{ph}$ (right).
Let us concentrate first on the isoscalar
mode (left).  The up-shift by the residual interaction is small for
the parametrizations with $m^*/m\approx 1$, in accordance with the
small values of $g$ or $G^{\vphu}_0$. In these cases, the $1ph$
energies represent already a good estimate of $E_1$ and the
theoretical $\ve^{\pi}_{ph}$ lie close to the experimental value
(faint dotted line).  Lower effective masses
increase
$\ve^{\pi}_{ph}$,
away from the wanted $E_1$, and need more residual interaction to
compensate. The impact of residual interaction is much larger for the
isovector modes (right), again in accordance with the much larger spin
coupling parameter $g'$. In that case, we also have the problem that
all theoretical $\ve^{\nu}_{ph}$ are much higher than the experimental
value of 5.84 MeV.

The upper right panel of
Figure \ref{fig:M1-refit} shows the $B(M1)$ strength integrated over
the vicinity of the upper M1-mode.
One observes a close relation between
$g'$ and the isovector $B(M1)$ value: An increase of $g'$
reduces the $B(M1)$ value.  This is due to the
increase of the ground state correlations
($Y$-components of the RPA transition amplitudes)
which decreases the transition probabilities in the magnetic case in
contrast to the electric case where the ground state correlations add
coherently. The parametrizations SkP$_{\mbss{m}}$ and
SLy5$_{\mbss{m}}$ behave slightly different because as mentioned
before in these parametrizations the $J^2$ terms are included.
These terms have a noticeable impact on the $B(M1)$ values that can be
estimated with the help of the single-particle part of the RPA
energy-weighted sum rule (EWSR) $m^{\mbsu{s.p.}}_1$. In the case of
the M1 excitations with the operator (\ref{def:qnp}) it has the form
\be
m^{\mbsu{s.p.}}_1 = \frac{1}{2}\,
\mbox{Tr}\,\bigl(\rho\,\bigl[\bigl[\,\bfbq,\,h\,\bigr],\cdot\,\bfbq\,\bigr]\bigr)
\label{ewsrsp}
\ee
(see Ref. \cite{Tselyaev_2011} for more details).  In our
self-consistent RPA calculations we obtain that this EWSR is fulfilled
within 0.2\% in the case of the Skyrme-EDF parametrizations without
the $J^2$ terms ($\eta^{\vphu}_J=0$).  In the case of the
SkP$_{\mbss{m}}$ and SLy5$_{\mbss{m}}$ parametrizations
($\eta^{\vphu}_J=1$), this EWSR is exceeded by 19 and 25\%,
respectively.

Generally, we see in the upper right panel of Figure
\ref{fig:M1-refit} that the theoretical isovector $B(M1)$ strengths,
even for the best parameter sets, are significantly larger than the
experimental values.  Here one has to bear in mind that the
experimental data in Figure \ref{fig:M1-refit} have been integrated
only up to 8.1 MeV. We know from previous beyond-RPA calculations
within the Landau-Migdal approach \cite{Kamerdzhiev_1993a} that the
theoretical strength is distributed by coupling to $2ph$ states up to
much higher energies.  Such spectral fragmentation is also seen in
data.  A recent $(p,p')$ experiment \cite{Poltoratska12,Birkhan_2016} reports a
summed $B(M1)$ = 20.5(1.3) $\mu^2_N$ when integrated up to 9 MeV, a
value which would fit nicely into the theoretical results of Figure
\ref{fig:M1-refit}. This situation reminds us at the case of the
Gamow-Teller resonance in $^{208}$Pb where only half of the sum-rule
strength was concentrated in one single strong resonance and the rest
was missing. Calculations within a $2ph$ model \cite{Drozdz_1990}
(where one of the authors was involved) predicted a long tail which
included the other half of the total strength. Ten years later the
predicted strength had been detected experimentally.
Thus, excess of the strength can be corrected in extended RPA models
including particle-phonon coupling that give also rise to a shift of
the RPA strength to higher energies.

So far, we have computed the $B(M1)$ strengths with the effective
M1 operator using the renormalization constants $\xi_s$ and $\xi_{\,l}$
as defined in Eq.~(\ref{xieff}).  This construction is designed to
account for correlation effects not included in the actual Hilbert
space. Thus the $\xi_s$ and $\xi_{\,l}$ can, in principle, be
different for the different models. This was exploited in the variant
SV-bas$_{\mbss{mx}}$ where $\xi_s$ was used tentatively as
further free parameter and the isovector $B(M1)$ strength as additional
data point.  The fitted renormalization constants for
SV-bas$_{\mbss{mx}}$ are $\xi_s=0.154$ while $\xi_{\,l}=0$ is chosen
in accordance with the condition (\ref{eqqe}).  The results in
Figure~\ref{fig:M1-refit} shows that this strategy allows to produce
better $B(M1)$ strength while maintaining the quality of the other
observables. Note that the changes in $\xi_s$ and $\xi_{\,l}$
  are, in fact, small which rather supports the original choices
  (\ref{xieff}). Anyway, this fit of renormalization constants should
  be considered as an exploration of still loose ends in
  modeling. Playing with these values needs yet to be supported by
  sound many-body theory.

As argued above, spin-orbit parameters have not only huge impact on M1
modes, but also on ground state properties.  Thus a dramatic change of
isovector spin-orbit coupling as implied in the re-tuned
parametrizations could have unwanted side effects on the quality
concerning the reproduction of ground state properties.
\begin{figure}[h!]
\centerline{\includegraphics[width=\linewidth]{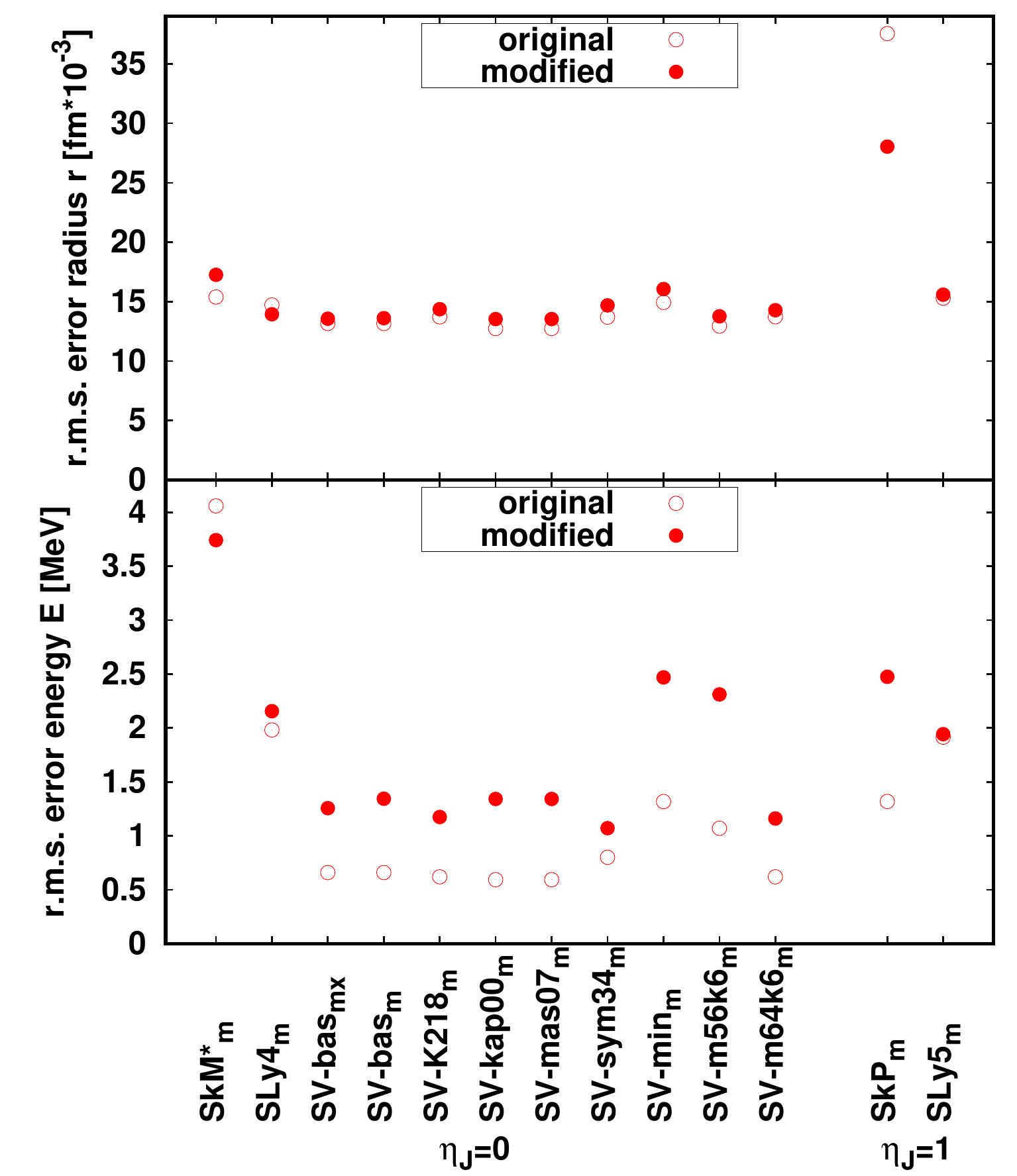}}
\caption{\label{fig:M1-refit-quality} Average quality of the
    re-tuned parametrizations quantified in terms of root-mean-square
    deviation of energy and charge radii taken over the set of
    spherical nuclei from \cite{Kluepfel_2009}.  }
\end{figure}
Figure~\ref{fig:M1-refit-quality} shows the performance of the
refitted parametrizations with respect to ground state
energy and charge radius.  The change of spin-orbit parameters leaves
the overall quality basically conserved.  There is no effect at all
for the radii. Energy reacts more sensitively which is little surprise
because pairing in semi-magic nuclei is highly sensitive to level
density which, of course, is influenced by spin-orbit splitting. Note
that particularly the more recent, well fitted parametrizations show a
loss of energy quality, fortunately in acceptable bounds.  Still, the
simple minded re-tuning strategy spoils somewhat the overall quality of
the parametrizations, the better the quality originally the larger the
loss.  Moreover, there are more subtle observables as pairing gaps and
isotopic shifts of radii. The latter are known to be sensitive to the
isovector spin-orbit term \cite{RF95}, for pairing gaps it is likely.
All this calls for more continued investigations, more systematic
fits, and correlation analysis \cite{Dob14a} to clearly work out the
impact of information from M1 modes on nuclear density functionals.

So far, we have discussed the properties of M1 modes in terms of two
energies and $B(M1)$ values.  Let us finally look again at the whole
spectral distribution as it was shown in Fig.~\ref{fig:m1rpa}.  The
results obtained with the freshly re-tuned parametrization
SV-bas$_{\mbss{mx}}$ agree, by construction, nicely with experimental
data. Comparison with the original SV-bas shows the gain. Similar
plots would be obtained when comparing original and re-tuned versions
of the other parametrizations.  But Fig.~\ref{fig:m1rpa} also points
toward the yet open problems with the upper M1 mode: First, the
strength is overestimated, and second, its spectral fragmentation is
not described at all.  Both problems are related to each other as
discussed above. The hope is that a beyond-RPA modeling within the
phonon-coupling model could deliver the missing pieces.

\section{Conclusions}

In the present paper, we investigate the dependence of the spin-dependent
part of the $ph$-interaction on the parameters of Skyrme
energy density functional (EDF).  This part is relevant for computing
magnetic excitation modes within the self-consistent
random-phase approximation (RPA). We considered here, in particular,
magnetic dipole (M1) modes in $^{208}$Pb as test case.  The M1 modes
are found depend crucially on the spin-orbit term and on the spin-spin
interaction. The latter has no influence on ground state properties
and generally only weak relations to natural-parity modes in even
nuclei and is thus open to adjustment. The spin-orbit term is to some
extend constrained by ground-state properties. However, we find that
ground states leave enough leeway in them to accommodate the
properties of M1 modes with only small sacrifices on the overall
quality of the ground state properties. We have tested that on a
variety of 12 published Skyrme EDFs.

In the analysis, we guide by the Landau-Migdal (LM) parameters
from the Theory of Finite Fermion Systems (TFFS) which are weak in the
isoscalar spin part and strongly repulsive in the isovector part.  The
re-tuned Skyrme EDFs deliver LM parameters in accordance with the TFFS.
The relations between the LM parameters and the parameters of the
Skyrme-EDF serve also for a quick first check of spin stability of the
chosen parameter set.

As open questions remain the fragmentation and the magnitude of the
isovector M1 resonance. Both are connected with more complex
configurations beyond RPA, e.g., the coupling to the low-lying phonons
(strong modes in each angular momentum channel). This, however,
requires that all relevant phonons, also in the magnetic channels, are
correctly described by RPA. The present survey is a first step toward
a proper description of magnetic excitations in the framework of
Skyrme-EDF and so paves the way to subsequent beyond-RPA calculations.

\bigskip


\begin{acknowledgements}
V.T. and N.L. acknowledge financial support from the Russian Science
Foundation (project No.  16-12-10155).
Research was carried out using computational resources provided
by Resource Center ``Computer Center of SPbU''.
\end{acknowledgements}


\appendix

\section{Local densities and currents}
\label{app:dens}

Let us introduce the isoscalar ($T=0$) and isovector ($T=1$) single-particle
density matrices
\bea
\rho^{\vphu}_T (\bfr,\sigma;\bfr',\sigma') &=&
\rho^{\vphu}_n (\bfr,\sigma;\bfr',\sigma')
\nonumber\\
&+& (-1)^T
\rho^{\vphu}_p (\bfr,\sigma;\bfr',\sigma')\,,
\label{def:rhom}
\eea
where
$\rho^{\vphu}_n(\bfr,\sigma;\bfr',\sigma')$ and
$\rho^{\vphu}_p(\bfr,\sigma;\bfr',\sigma')$ are the neutron's and proton's
density matrices.
The expressions for the local densities and currents entering Eq.~(\ref{Eden})
in terms of these matrices read
\bea
\rho^{\vphu}_T(\bfr) &=& \sum_{\sigma} \rho^{\vphu}_T (\bfr,\sigma;\bfr,\sigma)\,,
\label{def:rho}\\
\tau^{\vphu}_T(\bfr) &=& \sum_{\sigma} \nabla\cdot\nabla'\,
\rho^{\vphu}_T \bigl.(\bfr,\sigma;\bfr',\sigma)\bigr|_{\sbfr = \sbfrp}\,,
\label{def:tau}\\
\bfbj^{\vphu}_T(\bfr) &=& i \sum_{\sigma,\,\sigma'}\,
\bigl[ (\bfsigma)_{\sigma',\,\sigma} \times \nabla \bigr]\,
\rho^{\vphu}_T \bigl.(\bfr,\sigma;\bfr',\sigma')\bigr|_{\sbfr = \sbfrp}
\qquad
\label{def:bfbj}
\eea
for the time-even quantities and
\bea
\bfs^{\vphu}_{T}(\bfr) &=& \sum_{\sigma,\,\sigma'}\,
(\bfsigma)_{\sigma',\,\sigma}\;\rho^{\vphu}_T (\bfr,\sigma;\bfr,\sigma')\,,
\label{def:bfs}\\
\bfbt^{\vphu}_{T}(\bfr) &=& \sum_{\sigma,\,\sigma'}\,
(\bfsigma)_{\sigma',\,\sigma}\,\nabla\cdot\nabla'\,
\rho^{\vphu}_T \bigl.(\bfr,\sigma;\bfr',\sigma')\bigr|_{\sbfr = \sbfrp}\,,
\qquad
\label{def:bftau}\\
\bfj^{\vphu}_T(\bfr) &=& \frac{i}{2} \sum_{\sigma}\,\bigl( \nabla' - \nabla \bigr)\,
\rho^{\vphu}_T \bigl.(\bfr,\sigma;\bfr',\sigma)\bigr|_{\sbfr = \sbfrp}
\label{def:bfj}
\eea
for the time-odd quantities.

For the local densities $\tau_p(\bfr)$, $\tau_n(\bfr)$, and $\rho_p(\bfr)$
in Eqs. (\ref{Ekin}) and (\ref{ECoul}) we have
$\tau_p = (\tau^{\vphu}_0 - \tau^{\vphu}_1)/2$,
$\tau_n = (\tau^{\vphu}_0 + \tau^{\vphu}_1)/2$,
$\rho_p = (\rho^{\vphu}_0 - \rho^{\vphu}_1)/2$.

\section{Parameters of the Skyrme EDF}
\label{app:param}

The following equations establish the relation between
the $C$-constants in Eq.~(\ref{Eden}) and the parameters of the Skyrme force
$t_0$, $x_0$, $t_1$, $x_1$, $t_2$, $x_2$, $t_3$, $x_3$, $W_0$, and $x_W$
\begin{widetext}
\be
\begin{array}{llllll}
C_0^\rho &=& \ffrac{3}{8}  t_0\,,&
C_1^\rho &=& -\ffrac{1}{4}  t_0 (\half + x_0)\,,\\
C_0^{\rho,\alpha} &=& \ffrac{1}{16} t_3\,,&
C_1^{\rho,\alpha} &=& -\ffrac{1}{24} t_3 (\half+x_3)\,,\\
C_0^{\Delta \rho} &=&   -\ffrac{9}{64} t_1  +\ffrac{5}{64} t_2
            +\ffrac{1}{16}  t_2 x_2\,,&
C_1^{\Delta \rho} &=& \ffrac{1}{32} \big[ 3t_1 (\half+x_1)+t_2 (\half+x_2)\big]\,,\\
C_0^\tau &=& \ffrac{3}{16} t_1 + \ffrac{5}{16} t_2 + \ffrac{1}{4} t_2 x_2\,,&
C_1^\tau &=& -\ffrac{1}{8} \big[ t_1 (\half+x_1)-t_2 (\half+x_2)\big]\,,\\
C_0^{J} &=& \ffrac{1}{8} \big[t_1 \big( \half - x_1 \big)
- t_2 \big( \half + x_2 \big)\big] \eta^{\vphd}_J\,,&
C_1^{J} &=& \ffrac{1}{16} ( t_1 - t_2 )\,\eta^{\vphd}_J \,,\\
C_0^{\nabla J} &=& -\ffrac{1}{4}(2 + x_W) W_0\,,&
C_1^{\nabla J} &=& -\ffrac{1}{4} x_W W_0\,,\\
C_0^s &=& \ffrac{1}{2} C^{\vphu}_{\mbsu{N}}\,g
- \ffrac{1}{4} t_0 \big( \half - x_0 \big) \eta^{\vphd}_{s}\,,&
C_1^s &=& \ffrac{1}{2} C^{\vphu}_{\mbsu{N}}\,g'
- \ffrac{1}{8} t_0\,\eta^{\vphd}_{s}\,,\\
C_0^{s,\alpha} &=& - \ffrac{1}{24} t_3 (\half - x_3)\,\eta^{\vphd}_{s,\alpha}\,,&
C_1^{s,\alpha} &=& -\ffrac{1}{48} t_3\,\eta^{\vphd}_{s,\alpha}\,,\\
C_0^{\Delta s} &=& \ffrac{1}{32} \big[ 3 t_1 \big( \half - x_1 \big)
+ t_2 \big( \half + x_2 \big)\big] \eta^{\vphd}_{\Delta s}\,,\qquad\qquad&
C_1^{\Delta s} &=& \ffrac{1}{64} ( 3 t_1 + t_2 )\,\eta^{\vphd}_{\Delta s}\,.
\end{array}
\label{def:CT}
\ee
\end{widetext}
The formulas for the spin-orbit constants $C_T^{\nabla J}$ imply the parametrization
introduced in \cite{RF95,Sharma_1995} in which the spin-orbit term of the
interaction is treated in the Hartree approximation.
The parameters $W_0$ and $x_W$ are related
with the constants $b^{\vphu}_4$ and $b'_4$ of Ref.~\cite{RF95}
by the formulas:
$W_0=2b_4$, $x_W=b'_4/b^{\vphu}_4$.
The parameter $\eta^{\vphd}_J = 1$ if the $J^2$ terms are included in the Skyrme EDF
and $\eta^{\vphd}_J = 0$ if not.
In the standard parametrizations, the parameters $x_W$, $\eta^{\vphd}_{s}$,
$\eta^{\vphd}_{s,\alpha}$, and $\eta^{\vphd}_{\Delta s}$ are equal to 1,
the parameters $g$ and $g'$ are equal to 0.

\bibliographystyle{apsrev4-1}
\bibliography{TTT}
\end{document}